%% file: saez.tex
\documentclass[apJ]{emulateapj}
\usepackage{graphicx}
\usepackage{amssymb}
\usepackage{textcomp}

\newcommand{\cmsq}{\hbox{cm$^{-2}$}}

\newcommand{\flux}{\hbox{erg~cm$^{-2}$~s$^{-1}$ }}
\newcommand{\lumin}{\hbox{erg~s$^{-1}$}}

\newcommand{\cdfn}{\hbox{CDF-N}}
\newcommand{\cdfs}{\hbox{CDF-S}}
\newcommand{\chandra}{{\sl Chandra}}

\newcommand{\obs}{\hbox{0.5--8~keV}}
\newcommand{\rest}{\hbox{2--10~keV}}
\newcommand{\OF}{\hbox{observed-frame}}
\newcommand{\RF}{\hbox{rest-frame}}
\newcommand{\lnh}{\hbox{log~$N_{\rm H}$}}

\newcommand{\lLa}{\hbox{log~$L_{2-10}$}}

\newcommand{\fbin}{\hbox{$0.3\lesssim z \lesssim 0.96$}}
\newcommand{\sbin}{\hbox{$0.96\lesssim z \lesssim 1.5$}}
\newcommand{\tbin}{\hbox{$1.5\lesssim z \lesssim 3.3$}}
\newcommand{\Glx}{$\Gamma-L_{\rm X}$ }
\newcommand{\oOF}{\obs $ $ \OF}
\newcommand{\rRF}{\rest $ $ \RF}

\begin{document}


\title{CONFIRMATION OF A CORRELATION BETWEEN THE X-RAY LUMINOSITY
AND SPECTRAL SLOPE OF AGNs IN THE CHANDRA DEEP FIELDS.}

\author{C.~Saez \altaffilmark{1}, G.~Chartas \altaffilmark{1}, W.~N.~Brandt \altaffilmark{1},
B.~D.~Lehmer \altaffilmark{2}, F.~E.~Bauer \altaffilmark{3}, X.~Dai
\altaffilmark{4}, and G.~P.~Garmire \altaffilmark{1}}

\altaffiltext{1}{Astronomy and Astrophysics Department, Pennsylvania State
University, University Park, PA 16802, USA}

\altaffiltext{2}{Department of Physics, University of Durham, South  Road,
Durham, DH1 3LE, UK}

\altaffiltext{3}{Columbia Astrophysics Laboratory, Columbia University, New
York, NY 10027, USA}

\altaffiltext{4}{Department of Astronomy, The Ohio State University,
Columbus, OH 43210, USA}

\begin{abstract}
We present results from a statistical analysis of 173 bright radio-quiet
AGNs selected from the {\sl Chandra} Deep Field-North and {\sl Chandra}
Deep Field-South surveys (hereafter, CDFs) in the redshift range of
0.1~$\lesssim~z~\lesssim~$4. We find that the X-ray power-law photon index
($\Gamma$) of radio-quiet AGNs is correlated with their 2--10~keV
rest-frame X-ray luminosity ($L_X$) at the $>$~99.5\% confidence level
in two redshift bins, $0.3~\lesssim~z~\lesssim~0.96$, and
$1.5~\lesssim~z~\lesssim~3.3$ and is slightly less significant in the
redshift bin $0.96~\lesssim~z~\lesssim~1.5$. The X-ray spectral slope
steepens as the X-ray luminosity increases for AGNs in the luminosity range
$10^{42}$ to $10^{45}$ \lumin. Combining our results from the CDFs with
those from previous studies in the redshift range
$1.5~\lesssim~z~\lesssim~3.3$, we find that the $\Gamma-L_{\rm X}$
correlation has a null-hypothesis probability of 1.6 $\times10^{-9}$. We
investigate the redshift evolution of the correlation between the power-law
photon index  and the hard X-ray luminosity  and find that the slope and
offset of a linear fit to the correlation change significantly (at the
$>$~99.9\% confidence level) between redshift bins of
$0.3~\lesssim~z~\lesssim~0.96$ and $1.5~\lesssim~z~\lesssim~3.3$. We
explore physical scenarios explaining the origin of this correlation and
its possible evolution with redshift in the context of steady corona models
focusing on its dependency on variations of the properties of the hot
corona with redshift.
\end{abstract}

\section{INTRODUCTION}
It is important to extend the study of quasars to high redshifts in order
to understand their evolution and environments. A relevant conclusion from
modern studies is that the quasar luminosity function evolves positively
with redshift, having a comoving space density strongly peaked at $z
\approx 2$ \citep[e.g.,][]{Sch68,Boy87,War94}. More recent findings suggest
that the evolution of the space density of AGNs is strongly dependent on
X-ray luminosity ($L_{\rm X}$), with the peak space density of AGNs moving
to higher redshifts for more luminous AGNs \citep[e.g.,][]{Ued03,Has05}.

The X-ray band probes the innermost region of the central engines of AGNs.
The study of AGNs in the X-ray band provides important insights about their
central engines and the evolution of the AGN luminosity function. In most
AGNs, the observed X-ray continuum can be modeled using a power-law of the
form $N(E) = N_{0}(E/E_{0})^{-\Gamma}$, where $\Gamma$ is the photon index.
This power-law is attenuated by material in our Galaxy as well as material
intrinsic to the host galaxy. Several recent studies have centered on
estimating the distribution of intrinsic column densities ($N_{\rm H}$) and
the fraction of AGNs having $N_{\rm H} \gtrsim 10^{22}$ \cmsq.
Recent theoretical studies of AGNs \citep[e.g.,][]{Hop05} suggest that the
distribution of $N_{\rm H}$ is luminosity dependent; this is supported
observationally with the detection of an anti-correlation between the
obscuration fraction and luminosity
\citep[e.g.,][]{Stef03,Ued03,LaF05,Aky06}. We note, however, recent work by
\cite{Dwe06} reporting that the obscuration fraction may be independent of
luminosity. The dependence of the obscuration fraction on redshift is a
controversial issue. Some authors detect an increase of the obscuration
fraction with redshift \citep[e.g.,][]{LaF05,Tre06,Toz06}, while others do
not find any evidence for evolution \citep[e.g.,][]{Ued03,Aky06,Dwe06}.

A recent mini-survey of relatively high-redshift $(1.5 < z < 4)$
gravitationally lensed radio-quiet quasars (RQQs) observed with {\it
Chandra} and {\it XMM-Newton} \citep{Dai04} indicated a possible
correlation between the X-ray power-law photon index and X-ray luminosity.
This correlation, characterized by an increase of $\Gamma$ with $L_{\rm
X}$, was found for RQQs with 2--10~keV luminosities in the range $10^{43}$
to $10^{45}$ \lumin. Such a correlation is not found in nearby $z \lesssim
0.1$ quasars \citep[e.g.,][]{Geo00}. Several studies to date of
high-redshift quasars do not have large enough sample sizes in the
2--10~keV luminosity range $10^{43}$ to $10^{45}$ \lumin~ to place any
significant constraints on a possible $\Gamma-L_{\rm X}$ correlation
\cite[e.g.,][]{Ree00,Pag05}.

One of the concerns with the \cite{Dai04} analysis was that the limited
number of quasars in the sample, combined with the poor signal-to-noise
ratio (S/N) available for several of the observations and the relatively
large fraction of BAL quasars, may have led to problematic systematic
effects. The number of available lensed radio-quiet quasars used by
\cite{Dai04} was limited to a total of 25 sources, of which the brightest
11 had X-ray observations. In order to increase the size of the
high-redshift radio-quiet quasar sample, we have compiled a sample of 173
high-redshift AGNs with moderate-to-high S/N spectra available from the
\chandra\ Deep-Field-North and \chandra\ Deep-Field-South surveys
\citep[\cdfn$ $ and \cdfs, respectively; jointly CDFs;][]{Gia02,Ale03}.

The main scientific goal of this work is to constrain better the
$\Gamma-L_{\rm X}$ correlation found by Dai et al. (2004). The significant
increase in sample size allows us to place tighter constraints on the
significance of the correlation. We also test the correlation in narrower
redshift bands which will allow us to determine the epoch by which possible
changes in the average emission properties of AGNs occurred. Currently, the
two deepest X-ray surveys are the \cdfn$ $ and \cdfs$ $ with $\approx$2~Ms
and $\approx$1~Ms exposures, respectively. Both surveys cover
$\approx$300~arcmin$^2$ areas and target different regions of the sky
characterized by low Galactic column densities and an absence of bright
stars \citep{Gia02,Ale03}. The CDFs pointings have sufficient sensitivity
to detect the X-ray emission from AGNs with moderate luminosities ($L_{\rm
X} \approx10^{43}-10^{44}$ \lumin) out to $z\approx2-6$.

Radio-quiet AGNs (RQ AGNs) correspond to the majority of active
galaxies $(\sim90\%)$ that contain a central active nucleus
and show several differences in
their spectral properties compared to radio-loud AGNs. Radio-loud AGNs
have powerful sub-parsec jet-linked X-ray synchrotron self-Compton (SSC)
emission, which introduces an additional component to their spectra. As a
consequence RQ AGNs are observed to have, on average, steeper X-ray
power-laws than radio-loud AGNs \citep[e.g.,][]{Ree97}. We therefore have
chosen to exclude radio-loud AGNs from this study. Throughout this paper we
adopt a flat $\Lambda$-dominated universe with
$H_0$=70~km~s$^{-1}$Mpc$^{-1}$, $\Omega_\Lambda=0.7$, and $\Omega_M=0.3$.
The \chandra\ data were reduced using the CIAO version 3.3 software tools
provided by the {\sl Chandra} X-ray Center (CXC), and the spectral analysis
was performed using XSPEC version 12.

\begin{figure}
   \epsscale{1} \plotone{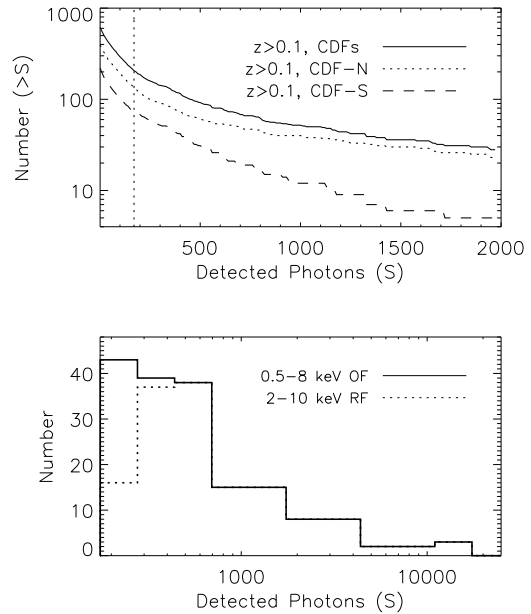}
        \centering
      \caption{(upper panel) Number of sources with more than $S$ photons (\oOF) versus
      $S$.  The thick line corresponds to all CDFs sources with measured redshifts of
      $z\gtrsim0.1$. The dotted line shows sources of the CDF-N survey with $z\gtrsim 0.1$, and the
     dashed line shows sources of the CDF-S survey with $z\gtrsim 0.1$.
     The vertical dotted line corresponds to $S$=170. Note that the sample used to
     generate this figure contains AGNs (both RQ and radio-loud AGNs), normal galaxies and starburst galaxies.
     (lower panel) Number of radio-quiet AGNs with $z\gtrsim 0.1$ vs. the number of photons ($S$; 0.5--8~keV) in their
     spectra. The solid line represents sources with fits performed in the \oOF $ $ band and the dotted
     line represents sources with fits performed in the \rRF $ $ band.}
     \label{cnts}
\end{figure}

\section{SAMPLE SELECTION}
Our sources were selected from the CDFs, currently the two deepest X-ray
surveys. The on-axis sensitivity limits for the \cdfn$ $ are $\approx 2.5
\times 10^{-17}$ \flux ($\mbox{0.5--2.0~keV}$) and $\approx 1.4 \times
10^{-16}$ \flux (2--8~keV). These limits are around two times more
sensitive than those for the \cdfs$ $ \citep{Gia02,Ale03}. The CDFs are
50--250 times more sensitive than previous X-ray surveys, detecting
$\approx 900$ point sources, of which $\approx 600$ are AGNs and galaxies
with measured redshifts \citep{Gia02,Ale03,Bar03,Zhe04}.

Spectroscopic and photometric redshifts were gathered from the literature
\citep{Cro01, Bar03, Ste03, Cow04, Mob04, Szo04, Wir04, Wol04, Zhe04,
Ale05, Col05, Fev05, Van06} and vetted to remove redshifts which did not
appear to belong to the most-likely optical counterpart to each X-ray
source. The latter was assessed by comparing the optical and X-ray images,
which were aligned to between 0\farcs12--0\farcs25; this notably affected
the redshifts from \citet{Zhe04}, where 47 ($\approx$14\%) of the redshifts
were rejected for being associated with an unlikely optical counterpart.
For the $\sim$40 faint sources still lacking redshift estimates, we used
the BPZ code \citep[Bayesian photometric redshift estimation;][]{Ben00} and
available photometry \citep{Arn01, Bar03, Gia04} to estimate crude
redshifts.

The selection criteria for our sample of RQ AGNs are (1) that the sources
are radio-quiet (see below), (2) that the redshifts of the sources are
greater than 0.1, and (3) that the total number of photons in the full band
(0.5--8 keV) is greater than $\sim$ 170 counts ($S\gtrsim170$) resulting in
moderate-to-high S/N spectra. The selection of a cut-off at $\sim$ 170
counts allows an accurate estimate of the photon index, which is not
possible for fainter sources \citep{Toz06}. Based on the condition
that $S\gtrsim170$, the on-axis flux limits of our sample in the full band
(\obs) in the CDF-N and CDF-S surveys are $\approx 1 \times 10^{-15}$ \flux
and $\approx 2 \times 10^{-15}$ \flux, respectively.

In Figure \ref{cnts} (upper panel), we show the cumulative distribution for
number of X-ray sources having more than $S$ counts (0.5-8~keV) for the
\cdfn$ $ and \cdfs. The CDFs contain 205 sources with more than 170
counts at $z>0.1$. Most of these sources are AGNs; however, in the
low-redshift regime of our sample $0.1 \lesssim z \lesssim 1.0$ we expect
only a small fraction of starburst and ``normal'' galaxies \citep{Bra05}.
Following the classification scheme discussed in
 \S4.1.1 of \cite{Baua04}, we found two starburst galaxies and one ``normal''
galaxy, which we remove leaving 202 AGNs in our sample.

Radio-loud AGNs were classified based on a radio-loudness parameter
$R\gtrsim10$ ($R=f_{\rm 5GHz}/f_{\rm B}$). To find these sources, we
matched the X-ray positions with radio sources using a matching radius of 2
arcsec. The flux-density at 5~GHz was obtained from the flux-density
at 1.4 GHz assuming a power law radio spectrum ($f_\nu \propto
\nu^{-\alpha_r}$), where $\alpha_r=0.8$ is a characteristic radio
spectral index of synchrotron radiation\footnote{In AGNs values of
$\alpha_r$ could be flatter than the adopted $\alpha_r=0.8$
\citep[e.g.,][]{Ric98,Mux05}, with measured standard deviations $\sim$1
\citep[e.g.,][]{Wad04}. We investigated how a flatter $\alpha_r$ may affect
our results and find that choosing a value of $\alpha_r=0.6$, for example,
to estimate $R$ will not change our sample of RQ AGNs, whereas, a value of
$\alpha_r=0.4$ will result in the exclusion of only two sources from our
sample (1\% of the entire sample) in order to satisfy $R\lesssim10$. We
conclude that our sample selection and results of our statistical analysis
are not significantly affected by values of the radio spectral index as low
as $\alpha_r = 0$.}.

The flux in the $B$ filter was
obtained from \cite{Bar03} for the \cdfn$ $ sources and from public-domain
tables of the GOODS and \hbox{COMBO-17} surveys for the \cdfs$ $ sources.
When searching the radio catalogs\footnote{The radio surveys of
\cite{Ric00} and \cite{Afo06} cover the entire \cdfn~ and \cdfs~ regions
respectively. More details of the \cdfs~ radio observations are found in
\cite{Nor06}.} provided by \cite{Ric00} for the \cdfn~  and \cite{Afo06}
for the \cdfs, we find that 29 ($\sim$14\%) out of the 202 X-ray detected
AGNs were radio-loud. This leaves 173 RQ AGNs which we use for our analysis
out of which 111 have spectroscopic redshifts.

\section{SPECTRAL EXTRACTION}

The X-ray spectra of the sources of the CDFs analyzed in our study were
extracted using the software routine {\sc acis\_extract} v3.94 (hereafter
{\sc ae};  Townsley et al. 2003; Broos et al. 2005), included in the Tools
for ACIS Real-time Analysis (TARA; Oct 20,  2005) software package.
\footnote{{TARA is available at \tt
http://www.astro.psu.edu/xray/docs/TARA/}} {\sc ae} is ideal for extracting
and analyzing the  spectra of large numbers of point and diffuse sources
observed with ACIS over multiple epochs. {\sc ae} calls procedures from
both CIAO (v3.3) and HEASOFT (v6.0.4) and uses calibration files that are
part of the CALDB v3.2.1 product  provided by the {\sl Chandra} X-ray
Center.

The $\approx$2~Ms CDF-N ($\approx$1~Ms CDF-S) observations are comprised of
20 (10) event files. The event files were corrected for charge transfer
inefficiency, bad columns, bad pixels, and cosmic ray afterglows. The event
files were also filtered for time intervals of acceptable  aspect solution
and background levels. A detailed description of the data reduction
procedures are presented in \citet{Ale03}. Background event files and
exposure maps were created by excluding circular regions centered on the
detected sources with radii that are a factor of 1.1 times larger than the
99\%  encircled energy radii of the point spread functions at $\sim$
1.49~keV. Source extraction regions were constructed to contain 90\% of the
PSF encircled energy derived from the CXC 1.4967~keV PSF libraries. There
were two exceptions to this procedure. First, for sources with greater than
1000 counts in the \citet{Ale03}  catalog we used extraction regions that
contained 99\% of the PSF encircled energy. Second, for sources with 90\%
encircled energy extraction regions  that overlapped we reduced the
extraction regions to avoid overlap. Local background extraction regions
were chosen as annuli centered on the source positions with inner radii
equal to that of  the source extraction regions and with outer radii
selected such that the background region contained at least 100  background
counts and had an area at least fours times that of the source region.

We note that in the current analysis we made no attempt to correct for
possible spectral variability over the few year period of the observations
of the CDFs. Spectra obtained are therefore time-averaged over the period
of the observations.

\section{SPECTRAL ANALYSIS}

Two energy bands were used to fit the {\sl Chandra} spectra: the \oOF $ $
and the \rRF. To obtain the maximum S/N we utilized the observed-frame
energy range of 0.5--8~keV. The lower energy bound was chosen because the
{\sl Chandra} effective area is not well calibrated below 0.5~keV, and the
upper energy bound was chosen because the S/N decreases greatly above this
energy for most of the sources in the sample. One advantage of using the
same observed-frame energy range for every object is that the same
systematic instrumental uncertainties apply to every fit. Since most of the
detected spectrum is used in the analysis, the S/N is higher than for cases
where restricted energy ranges were used.

To test how the $\Gamma-L_{\rm X}$ correlation might be affected by
absorption and possible contamination from other emission processes, we
also fitted the spectra in the rest-frame energy range of~
$\mbox{2--10~keV}$. This range was selected to avoid possible contamination
from soft-excess emission that is often detected in AGNs below rest-frame
energies of $\sim$ 1~keV. The selection of the \rRF $ $ band also aids in
reducing the effects of X-ray absorption. For example, assuming a source
with a power-law spectrum of $\Gamma$=1.7, $z$=1, $N_{\rm
H}$$\sim$10$^{22}$\cmsq, and solar abundances the fraction of absorbed
photons is 30\% in the \oOF $ $ band and 9\% in the \rRF band. The \rRF $ $
also minimizes possible contamination from Compton-reflection emission from
circumnuclear material that is thought to peak at a rest-frame energy of
about 20~keV.
In general, 2--10~keV rest-frame spectra have fewer counts than \oOF $ $
spectra.
For fits performed in the 2--10~keV rest-frame band, we selected sources
with more than 170 counts in this band, leaving a sub-sample of 144 RQ
AGNs.

The total number of photon counts per source ($S$) with energies in the
\oOF $ $ band lies in the range $\mbox{170--13000}$. In Figure \ref{cnts}
(lower panel) we present the number of $z>0.1$ radio-quiet AGNs in our
sample versus the number of photons with energies in the 0.5--8~keV
observed-frame band. The solid line applies to sources with spectral fits
performed in the \oOF, and the dotted line applies to sources with spectral
fits performed in the \rRF.
The mean logarithm of $S$ for sources with spectral fits performed in the
\oOF $ $ is $\langle$log~$S \rangle$=2.74 with a standard deviation of
$\sigma\simeq 0.42$. The mean logarithm of $S$ for sources with spectral
fits performed in the \rRF $ $ is $\langle$log~$S \rangle$=2.83 with a
standard deviation of $\sigma\simeq 0.41$. Based on the fact that our
sample contains sources with relatively low counts, we used the
$C$-statistic \citep{Cas79} to fit spectra as adopted in a similar study
presented in \cite{Toz06}. In this study, the authors concluded that the
$C$-statistic is more accurate than the $\chi^2$-statistic in estimating
the spectral parameters of AGNs with low-count spectra ($\sim$ 100 counts);
similar arguments are presented in \cite{Nou89}. We also performed spectral
fits in the 0.5--8~keV observed-frame band using the $\chi^2$-statistic,
with a grouping of 10 counts per bin.
The sole purpose of using the $\chi^2$-statistic was to apply the $F$-test
to assess the use of more complex spectral models.

\input{t1.tex}

For the CDF-S and CDF-N sources of our sample, we assumed Galactic column
densities of $8.8 \times10^{19}$ \cmsq $ $ \citep{Sta92} and
$1.3\times10^{20}$ \cmsq $ $ \citep{Loc04}, respectively. The spectral
analysis was performed using XSPEC version 12. The default spectral model
used is a power law (PL; POW) with Galactic absorption (WABS). Additional
model components were added to the default model in cases where the
$F$-test showed an improvement in the fit at the 95\% confidence level
(0.5--8~keV observed-frame) when these additional components were used. We
refer to models comprised of the default model plus additional model
components as alternative models. Alternative models included an
absorbed-power-law model (APL) at the redshift of the source (WABS ZWABS
POW), an ionized-absorbed-power-law model (IAPL) (WABS ABSORI POW), a
partial-absorbed-power-law model (PAPL) (WABS ZPCFABS POW) and/or models
that included an iron line (PL+EL;APL+EL)(WABS ZGAUSS POW; WABS ZWABS
ZGAUSS POW). We also considered models that contained a Compton-reflection
component (PEXRAV), but we did not find any improvement in the fits using
such models. Our finding, that none of the sources in our sample require a
Compton-reflection component, is in agreement with \cite{Toz06} who find
that only 14 out of 321 CDF-S sources require Compton-reflection
components. We note that none of these 14 sources are part of our sample
mostly because they contain less than 170~counts. Even though we do not
detect a significant reflection-component in our spectral fits, an
unaccounted reflection-component could still be affecting the estimation of
$\Gamma$ (see \S5.2.6 for details). In Table \ref{tab:mode}, we list the
number of sources from our sample (the entire sample contains 173 RQ AGNs)
fit with a particular spectral model. From Table \ref{tab:mode}, we notice
that based on the $F$-test there are 92 sources with detected absorption
($\sim$53 \% of the whole sample) and 7 cases with detected iron lines
($\sim$4 \% of the whole sample).

\input{t2.tex}

The spectral-fitting results are presented in Table \ref{tab:CDFs} for
fits performed in the 0.5--8~keV observed-frame band (173 RQ AGNs), and the
fits performed in the 2--10~keV rest-frame band (144 RQ AGNs). In
Table \ref{tab:CDFs} we provide the photon index $\Gamma$ (errors at the
90\% confidence level), the intrinsic column density $N_{\rm H}$ (errors at
the 90\% confidence level) in units of $10^{22}$ \cmsq, and the logarithm
of the hard X-ray luminosity in the rest-frame 2--10~keV band in units of
\lumin (hereafter referred to as $L_{2-10}$). Table \ref{tab:CDFs}
also includes the X-ray identification of the sources based on their RA
and DEC positions, the photon counts in the fitted range, the number of
degrees of freedom, and the values of the $C$-statistic. The last two
quantities provide an estimate of the quality of the fits.

\begin{figure}
  \epsscale{1} \plotone{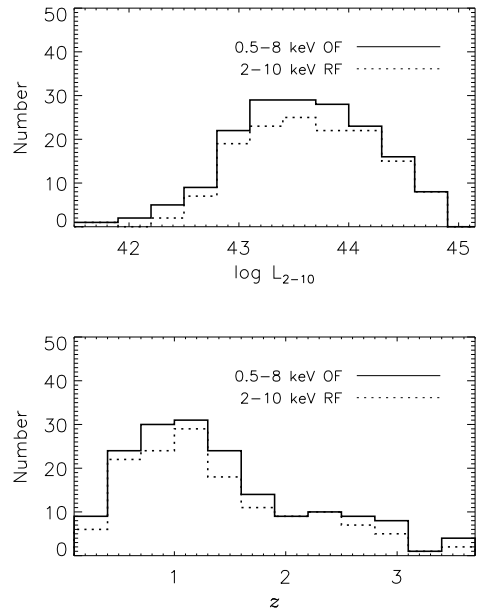}
        \centering
      \caption{Number of sources vs. luminosity (upper panel) and redshift (lower panel). The solid line represents sources with fits performed in the \oOF $ $ band and the dotted line represents sources with fits performed in the \rRF $ $ band.}
        \label{hist}
\end{figure}

In Figure \ref{hist}, we show the distributions of 2--10~keV luminosities
(upper panel) and redshifts (lower panel) of the sources in our sample with
fits performed in the \oOF $ $ band (solid line; 173 RQ AGNs) and the \rRF
$ $ band (dotted line; 144 RQ AGNs). The luminosities of the sources in our
sample cover the range $3\times10^{41}-6\times10^{44}$ \lumin$ $ where the
lower limit is mostly determined by the sensitivity limit of the CDF-N
survey, while the upper limit is a statistical consequence of the fact that
luminous AGNs ($L_{2-10}\gtrsim 10^{45}$ \lumin) are less numerous than
lower luminosity AGNs
\citep[see e.g.,][]{Bra05}. The mean redshift and mean logarithmic X-ray
luminosity of the sources with fits performed in the \oOF $ $ band are
$\langle z \rangle\simeq 1.41 $ and $\langle$\lLa$\rangle\simeq 43.6 $,
respectively. The mean redshift and mean logarithmic X-ray luminosity of
the sources with fits performed in the \rRF $ $ band are $\langle z
\rangle\simeq 1.38 $ and $\langle$\lLa$\rangle\simeq 43.6 $, respectively.

\begin{figure}
   \epsscale{1} \plotone{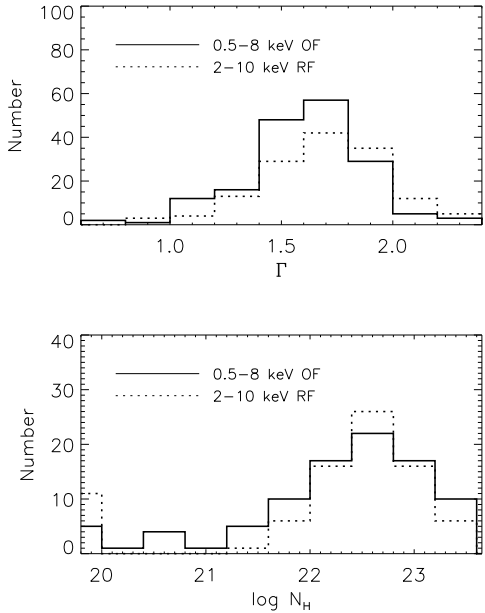}
        \centering
      \caption{Number of sources vs. photon index (upper panel) and column density (lower
panel). In the lower panel we show only the sources with
a significant detection of absorption in their spectra. 
The solid line represents sources with fits performed in the \oOF $ $ band
and the dotted line represents sources with fits performed in the \rRF $ $
band.}
        \label{dist}
\end{figure}

In Figure \ref{dist} (upper panel), we show the distributions of the photon
indices
of the sources with fits performed in the \oOF $ $ band (solid line) and
the \rRF $ $ band (dotted line). We find the mean photon indices and their
standard deviations for fits performed in the \oOF $ $ and \rRF $ $ bands
to be $\langle \Gamma \rangle \simeq 1.60$ $\pm$ 0.27 and $\langle \Gamma
\rangle \simeq 1.70$ $\pm$ 0.29, respectively.
In Figure \ref{dist} (lower panel), we show the distributions of the
intrinsic column densities of sources with significant absorption (only
sources where the $F$-test indicated significant intrinsic absorption at
the $>$95\% confidence level are included in the distributions); sources
with fits performed in the \oOF $ $ band are indicated with the solid line
(92/173), and sources with fits performed in the \rRF $ $ band are
indicated with the dotted line (82/144). In the two fitted energy ranges,
the peak of intrinsic column density distribution is \lnh~$\sim 22.6$, and
there is a fraction of $\sim$40\% sources from the total sample having
\lnh~$>22$. We note that there are likely systematic errors on these column
density estimates due to unmodeled absorption complexity.
These parameter values and distributions are in agreement with those found
in \cite{Toz06} and \cite{Dwe06}.

\begin{figure}
   \epsscale{1} \plotone{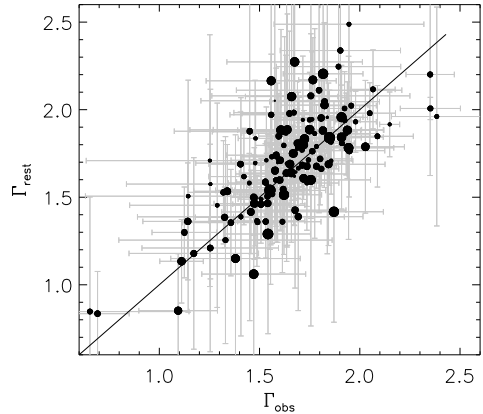}
        \centering
      \caption{Spectral index for fits performed in the  2--10~keV rest-frame band  ($\Gamma_{\rm rest}$)
      versus spectral index for fits performed in the 0.5--8~keV observed-frame band ($\Gamma_{\rm obs}$).
              The size of each symbol increases with redshift. The solid line represents the
               case of $\mbox{$\Gamma_{\rm obs} = \Gamma_{\rm rest}$}$. Notice that values of $\Gamma_{\rm rest}$ and
                $\Gamma_{\rm obs}$ can be found in Table 2.}
        \label{Gams}
\end{figure}

In Figure \ref{Gams}, we present a diagram comparing estimates of $\Gamma$
obtained from fits performed in the 2--10~keV rest-frame ($\Gamma_{\rm
rest}$) and fits in the \oOF $ $ ($\Gamma_{\rm obs}$). The size of each
symbol in Figure \ref{Gams} increases with redshift. Deviations from the
straight line ($\Gamma_{\rm obs}$=$\Gamma_{\rm rest}$) are most likely
statistical in nature, however, a few may be associated with the effects of
intrinsic absorption, soft excesses, non-detected spectral lines, and
Compton reflection.

In general, the agreement between $\Gamma_{\rm rest}$ and $\Gamma_{\rm
obs}$ is good; this is first quantified by a high Pearson linear
correlation coefficient ($\sim$0.73) and a very low null hypothesis
probability ($\sim 4.8 \times 10^{-25}$). Secondly, this agreement is
quantified by testing whether the linear relation between $\Gamma_{\rm
rest}$ and $\Gamma_{\rm obs}$ is consistent with $\Gamma_{\rm
rest}=\Gamma_{\rm obs}$. To verify the later we performed a $\chi^2$ fit to
the data assuming $\Gamma_{\rm rest}=\alpha \Gamma_{\rm obs}$, where
$\alpha$ was a free parameter.\footnote{A $\chi^2$ fit using the relation
y=$\alpha$x to model some bivariate sample ($x_i$, $y_i$) with errors in
both variables ($\sigma_{xi}$, $\sigma_{yi}$) is obtained by minimizing
$\chi^2=\displaystyle \sum^{}_{i}\frac{(y_i-\alpha
x_i)^2}{\sigma_{yi}^2+\alpha^2\sigma_{xi}^2}$} We considered the errors in
both variables $\Gamma_{\rm rest}$ and $\Gamma_{\rm obs}$ when performing
the least-squares fit. We obtained $\alpha=0.996\pm0.08$ (error at the 68\%
confidence level) with $\chi^2=99.3$ for 143 degrees of freedom (dof). As a
basic check for the luminosity dependence of the linear relation between
$\Gamma_{\rm rest}$ and $\Gamma_{\rm obs}$ we performed $\chi^2$ fits of
the model $\Gamma_{\rm rest}=\alpha \Gamma_{\rm obs}$  to sources with
$\lLa\lesssim43.6$ and sources with $\lLa\gtrsim43.6$. We obtained
$\alpha=0.995\pm0.011$ ($\chi^2=43.8$; dof=70) for sources with
$\lLa\lesssim43.6$ and $\alpha=0.997\pm0.010$ ($\chi^2=55.6$; dof=72) for
sources with $\lLa\gtrsim43.6$.

\begin{figure}
   \epsscale{1} \plotone{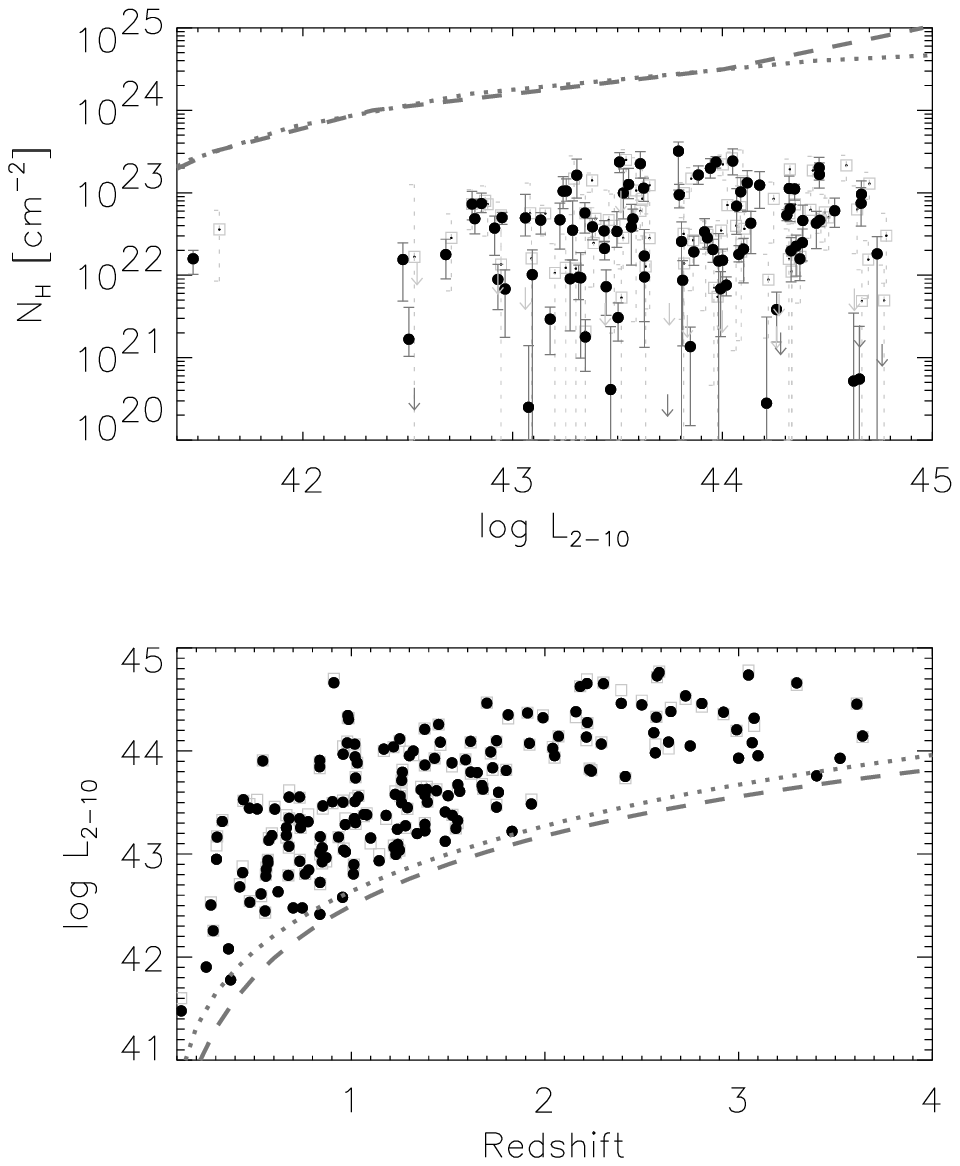}
        \centering
    \caption{Estimated best-fit column densities
versus 2--10~keV luminosities (upper panel), and 2--10~keV luminosities
versus redshifts (lower panel) of the $z > 0.1$ RQ AGNs. Filled circles
represent sources with fits performed in the \oOF $ $ band and open squares
represent sources with fits performed in the \rRF $ $ band. In the upper
panel, the two lines indicate the maximum column density that can be
detected for a source with $0.1 \leq z \leq 4.0$, and a total of 170 counts
in the \oOF $ $ (dashed) and \rRF $ $ (dotted). In the lower panel the
lines indicate the minimum luminosity required for the detection of a
source as a function of redshift. A total of 170 counts in the \oOF $ $
band (dashed) or a total of 170 counts in the \rRF $ $ band (dotted) is
assumed. For the limits shown in this figure it is assumed that the source
is detected at the ACIS-I aim-point with an exposure time of
$\approx$2~Ms.}
        \label{sele}
\end{figure}

\section{RESULTS AND DISCUSSION}

\subsection{Selection Effects.}
We begin this section by describing several selection effects that are
intrinsic to our sample. Having an exposure of $\approx2$ times greater
than the CDF-S survey, the \cdfn$ $ survey drives the sensitivity limits of
our sources, so the following discussion will be focused on this survey.

Figure \ref{sele} shows the estimated best-fit column density versus the
2--10~keV luminosity (upper panel), and the 2--10~keV luminosity versus
redshift (lower panel), for our $z > 0.1$ RQ AGNs. Spectral fits performed
in the 0.5--8~keV observed-frame band and \rest $ $ rest-frame band are
shown with filled circles and open squares respectively. In the upper panel
of Figure \ref{sele}, the dashed line shows the maximum column density that
can be found for a source with $0.1 \leq z \leq 4.0$
assuming the source is detected at the ACIS-I aim-point with an exposure
time of $\approx$2~Ms \citep{Ale03} and a total of 170 counts in the \oOF $
$ band (dashed line) and \rRF $ $ band (dotted line). Each point on these
curves is obtained by fixing $N_{\rm H}$ and finding the minimum luminosity
that can be obtained with
$0.1 \leq z \leq 4.0$ assuming a source with $\Gamma=1.6$, Galactic column
density of 1.3$\times$10$^{20}$\cmsq $ $ and 170 counts in each fitted
energy range. For low-luminosity sources ($L_{2-10}\lesssim 10^{42}$
\lumin), the threshold column density is $\approx$10$^{24}$~\cmsq, which
increases by a factor of $\sim$10 for higher luminosity sources
($10^{44}-10^{45}$ \lumin). The maximum column density observed at a
specific luminosity is set by AGNs with $z=0.1$ in most of the observed
luminosity range; however, for luminous sources ($L_{2-10} \gtrsim
10^{44}$~\lumin) higher redshift AGNs ($z$ $\sim$ 1--4) establish the limit
in $N_H$ because these are less affected by absorption.

In the lower panel of Figure \ref{sele}, the curves indicate the minimum
luminosity required for the detection of a source as a function of
redshift. We have assumed a source free of intrinsic absorption, positioned
at the ACIS-I aim-point with an exposure time of $\approx$2~Ms,
$\Gamma=1.6$, galactic column density of 1.3$\times$10$^{20}$\cmsq $ $ and
with 170 photon counts in the \oOF $ $ (dashed line) or 170 counts in the
\rRF $ $ (dotted line). The threshold luminosity is $\approx$10$^{42}$
\lumin$ $ for $z\approx0.5$ and $\approx$3$\times10^{43}$ \lumin$ $ for
$z\approx2.5$. The dashed curve in Figure \ref{sele} is obtained by
assuming no intrinsic absorption; however, the presence of $N_{\rm H}$,
which might be evolving \citep[e.g.,][]{LaF05,Tre06,Toz06}, could be
increasing the observed threshold luminosity.

\input{t3.tex}

\begin{figure}
   \epsscale{1} \plotone{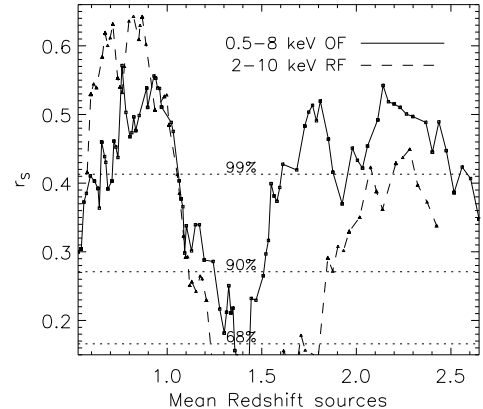}
        \centering
      \caption{Spearman correlation coefficients of the $\Gamma-L_{2-10}$
      relation as a function of the mean redshift within each sub-sample.
      Each sub-sample contains 38 RQ AGNs. The solid line corresponds to the fits performed
      in the 0.5--8~keV observed-frame band (Table \ref{tab:CDFs}). The dashed-line corresponds to the fits performed
      in the 2--10~keV rest-frame band (Table \ref{tab:CDFs}). The dotted lines correspond to three different levels of
      significance (68\%, 90\% and 99\%), assuming 38 independent measurements of $\Gamma$ vs. $L_{2-10}$.}
        \label{maxc}
\end{figure}

\begin{figure}
   \epsscale{1} \plotone{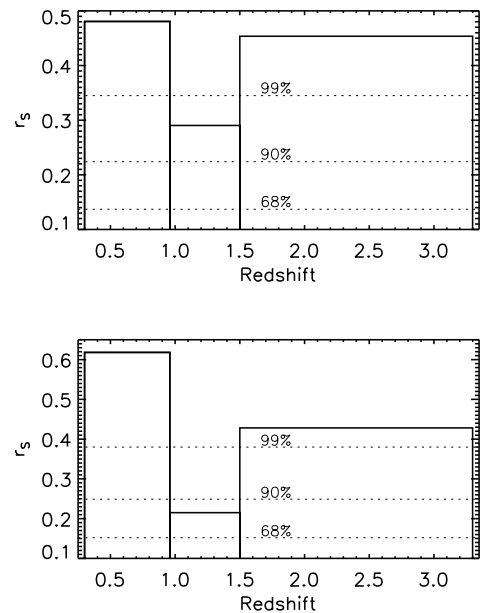}
        \centering
      \caption{Spearman correlation coefficients of the $\Gamma-L_{2-10}$
      relation as a function of redshift for the RQ AGNs within each independent redshift bin.
      In the upper panel each redshift bin contains $\sim$55 RQ AGNs and the fits were performed
      in the 0.5--8~keV observed-frame band (Table \ref{tab:CDFs}).
      In the lower panel each redshift bin contains $\sim$45 RQ AGNs, and the fits were performed
      in the 2--10~keV rest-frame band (Table \ref{tab:CDFs}). The dotted lines correspond to three different levels of
      significance (68\%, 90\% and 99\%); these are obtained assuming 55 sources in the upper panel and 45 sources in the
      lower panel.}
          \label{barc}
\end{figure}

\subsection{Luminosity and Photon index}

One of the goals of this work is to examine a possible correlation between
$L_{\rm X}$ and $\Gamma$ in a sample of RQ AGNs which was previously
reported by \cite{Dai04}. To improve on the Dai et al. analysis we
significantly increased the sample size using the CDFs, considered a larger
redshift range, and used X-ray spectra that contained more than 170 counts
in the full band (0.5--8~keV). The results of this analysis is shown in the
following sections. For the following analysis we use the X-ray luminosity
in the 2--10~keV rest-frame band ($L_{2-10}$).

\begin{figure}
   \epsscale{1} \plotone{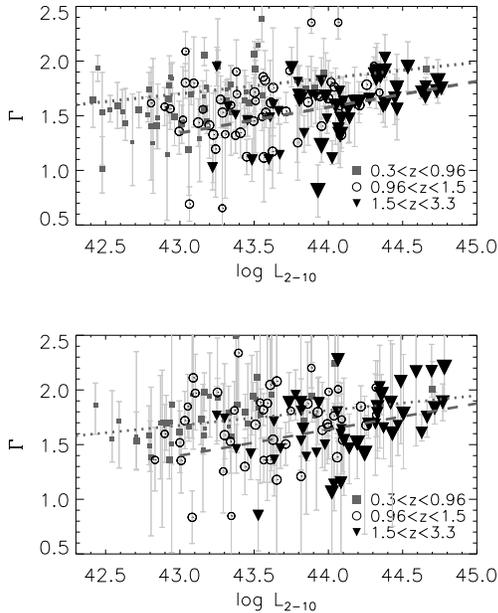}
        \centering
      \caption{$\Gamma$ versus 2--10~keV luminosity of radio-quiet AGNs in the redshift range of
        $\mbox{$0.3 \lesssim z \lesssim 3.3$}$.
       In the upper panel we show sources with fits performed in the \oOF $ $ band.
       In the lower panel we show sources with fits performed in the \rRF $ $ band.
       The symbol size increases with redshift.
       Filled squares are sources with $0.3 \lesssim z \lesssim 0.96$, open circles are sources with $0.96 \lesssim z \lesssim 1.5$,
       and filled triangles are sources with $1.5 \lesssim z \lesssim 3.3$. The dotted line indicates the least-squares fit to sources having $0.3\lesssim z \lesssim 0.96$. The dashed line shows the least-squares fit to sources having $1.5\lesssim z \lesssim 3.3$.}
        \label{alll}
\end{figure}

\subsubsection{Possible evolution of the strength and significance of the $\Gamma-L_{\rm X}$ correlation}

As a first approach, we searched for a $\Gamma $ -- $L_{\rm X}$ correlation
as a function of redshift selecting sub-samples ordered in redshift.
We used sub-samples containing 38 sources for fits performed in the
0.5--8~keV observed-frame band and the 2--10 keV rest-frame band (Table
\ref{tab:CDFs}). We calculated the mean redshift of each sub-sample and
computed the Spearman rank correlation coefficient and significance of the
correlation between $\Gamma $ and $L_{\rm 2-10}$. This process was repeated
by shifting the sampling window across the entire observed redshift range.
Figure~\ref{maxc} shows the values of the significance of the correlations
and the Spearman correlation coefficients of the $\Gamma-L_{2-10}$ relation
as a function of the mean redshift of the sources within each sub-sample,
using the best-fit parameters from Table~\ref{tab:CDFs}. The solid
line in Figure~\ref{maxc} corresponds to sources fitted in the 0.5--8~keV
observed-frame band and the dashed line corresponds to those fitted in the
2--10~keV rest-frame band. We notice that the correlation has two
significant peaks ($\sim$99\%) in both energy bands fitted, one with a mean
redshift of $\sim 0.7$, and the other with a mean redshift of $\sim 2.2$.

As a second approach, we selected three independent redshift bins covering
the redshift range $0.3<z<3.3$. The high redshift bin (\tbin) was chosen to
match the redshift range where \cite{Dai04} found the \Glx correlation
while the other two redshift bins (\fbin $ $ and \sbin) were selected to
obtain independent redshift bins with comparable numbers of sources within
them. Each redshift bin contained $\sim 55$ sources in the 0.5--8~keV
observed-frame band and $\sim 45$ sources in the 2--10~keV rest-frame band.
In Figure~\ref{barc} and Table \ref{tab:corr} we show the Spearman
rank correlation coefficient and the significance of the Spearman
correlation coefficient in each bin. The upper panel of Figure \ref{barc}
corresponds to fits performed in the 0.5--8~keV observed-frame and the
lower panel to fits performed in the 2--10 keV $\mbox{rest-frame}$. The
height of each bar is the significance of the $\Gamma-L_{2-10}$ Spearman
correlation. The correlation for fits performed in the 0.5--8~keV
observed-frame is significant for the three redshift bins; however, we find
a slight decrease in the strength and significance in the second redshift
bin for fits performed in the \rest $ $ rest-frame. The significance of the
correlation in the first and third redshift bins is $>$99.5\% for both
fitting ranges. A significant expansion of our sample made by incorporating
additional deep AGN surveys will be required to confirm the possible
decrease of the strength of the correlation in the second redshift bin.

\begin{figure}
   \epsscale{1} \plotone{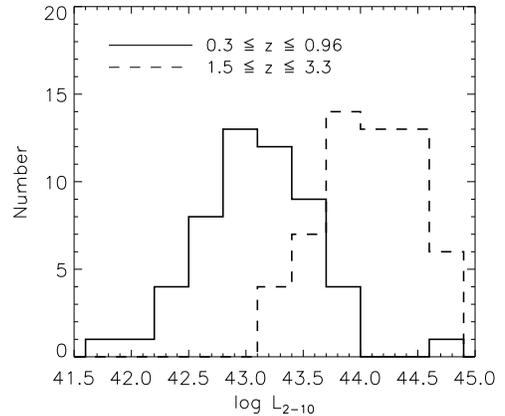}
        \centering
      \caption{2--10~keV rest-frame luminosity ($L_{2-10}$) distributions
      for radio-quiet AGNs with $0.3 \lesssim z \lesssim 1.5$
      (thick line) and $1.5 \lesssim z \lesssim 3.3$ (dashed line). The fits are performed in the \oOF $ $  band.}
        \label{lbin}
\end{figure}

In Figure \ref{alll}, we plot $\Gamma$ vs. $L_{2-10}$ for sources in each
redshift bin of Figure \ref{barc}, for fits performed in the 0.5--8~keV
observed-frame (upper panel) and the 2--10~keV rest-frame (lower panel).
Sources in our sample with redshifts in the range $0.3 \lesssim z \lesssim
0.96$ have a lower mean luminosity of $ \langle$\lLa$\rangle \sim 43.1$
($\sigma_{{\rm log}L_{2-10}} \sim 0.5$) than sources with redshifts in the
range
 $1.5 \lesssim z \lesssim 3.3$ which have  a mean luminosity of
$\langle$\lLa$\rangle \sim 44.1$ ($\sigma_{{\rm log}L_{2-10}} \sim 0.4$).
The luminosity distributions for sources in the redshift bins $0.3 \lesssim
z \lesssim 0.96$ and $1.5 \lesssim z \lesssim 3.3$ are shown in Figure
\ref{lbin}. We notice that the peak of the distribution of the sources in
the high-redshift bin is significantly higher in luminosity than the peak
of the distribution of the sources in the low-redshift bin. This shift in
luminosity distributions is mainly a selection effect (see
Figure~\ref{sele}) combined with the fact that luminous sources are more
numerous at high redshift \citep{Ued03,Has05}. The Spearman correlation
index of the $\Gamma-L_{2-10}$ data for the whole sample (173 RQ AGN) is
$\sim$0.24 (99.8\% significance) and $\sim$0.16 (94.1\% of significance)
for the \oOF $ $ and the \rRF $ $ spectral fits. These correlation
coefficients are significantly lower than those found in the  $0.3 \lesssim
z \lesssim 0.96$ and $1.5 \lesssim z \lesssim 3.3$ redshift bins
(see Table \ref{tab:corr} for more details).

In Table \ref{tab:spec}, we show the results of a test of the
$\Gamma-L_{\rm X}$ correlation using only sources with spectroscopic
redshifts. We find that the $\Gamma-L_{\rm X}$ correlation of the
sub-sample of sources with spectroscopic redshifts is significant in the
three redshift bins; however, as indicated in Table \ref{tab:spec}, this
sub-sample includes a larger fraction of type 1 AGNs and contains more
sources with \lnh $\lesssim$ 22 than that of the whole sample. In addition,
the size of this sub-sample is significantly smaller than the whole sample.
We caution that the strengths of the correlations provided by the
non-parametric tests used in our analysis of sub-samples containing a small
number of sources $ N $ with the present uncertainties in the photon
indices may be inaccurate since the variance of the Spearman correlation
coefficient is $\sigma^2$=$\frac{1}{N-1}$. Photometric redshifts are
subject to larger errors than spectroscopic ones and for sources with $z >
1$ the error is approximately given by \mbox{$\Delta z/(1+z)=0.05$}
\citep[e.g,][]{Coh00}. In our analysis, the uncertainties in the redshifts
will mainly affect the estimation of the X-ray luminosities. For example, a
source at $z \sim 2$ will have an uncertainty in the estimated luminosity
of $\Delta L_{\rm X}/L_{\rm X}\sim 0.3$. This level of uncertainty will not
significantly affect our results since our study involves estimating
changes in the photon index over two orders of magnitude in X-ray
luminosity.

In the following three sections, we focus on sources in the first bin ($0.3
\lesssim z \lesssim 0.96$) and the third bin ($1.5 \lesssim z \lesssim
3.3$)  and test the sensitivity of the $\Gamma-L_{\rm X}$ correlation to
the possible presence of intrinsic absorption and Compton reflection in the
spectra of the sources.

\input{t4.tex}

\begin{figure}
   \epsscale{1} \plotone{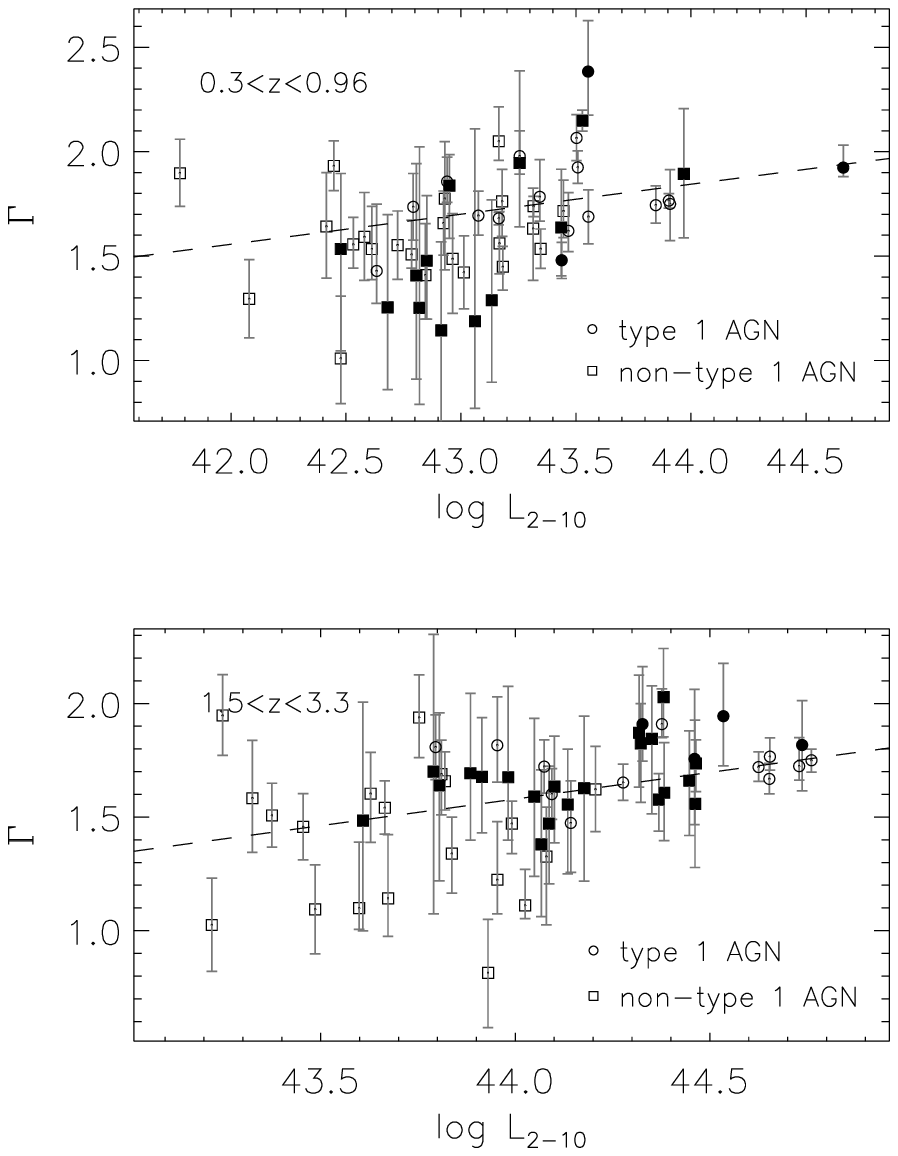}
        \centering
      \caption{$\Gamma$ versus 2--10~keV luminosity of radio-quiet AGNs with
      \fbin~ (upper panel) and with \tbin~ (lower panel).
      The values of the X-ray luminosities and spectral indices were obtained by fitting the spectra in the
      observed-frame energy range of \obs~ (see Table \ref{tab:CDFs}). The dashed lines indicate
      linear fits to the data using the least-squares method. The open symbols correspond
      to sources having $\mbox{\lnh~$\lesssim$~22}$, and the filled symbols are sources with \lnh~$>$~22.
      Circles correspond to type 1 AGNs and squares to non-type 1 AGNs.}
        \label{pmao}
\end{figure}

\begin{figure}
   \epsscale{1} \plotone{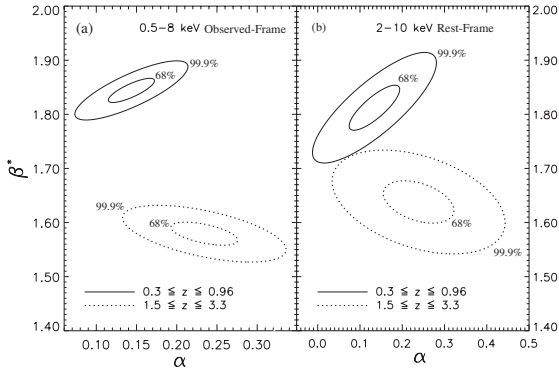}
        \centering
      \caption{68\% and 99.9\% confidence contours of the slope $\alpha$ and offset $\beta^*$ of the $\Gamma-L_{\rm X}$ correlation
for AGNs in the $0.3 < z < 0.96$ (solid contours) and $1.5 < z < 3.3$
(dotted contours) redshift ranges. The parameters $\alpha$ and $\beta^*$
were derived from fits of the linear model $\Gamma~=~\alpha~ {\rm log~}
\frac{L_{2-10}}{(10^{44}~{\rm ergs/s)}}~+~\beta^*$. The fits were performed
in the 0.5--8~keV observed-frame band (a) and in the 2--10~keV rest-frame
band (b). The confidence contours indicate that the parameters of the
linear fit to the $\Gamma-L_{\rm X}$  correlation differ at the $>$~99.9\%
confidence level between the  $0.3 < z < 0.96$ and $1.5 < z < 3.3$ redshift
ranges.}
        \label{conf}
\end{figure}

\subsubsection{Possible evolution of the slope and offset of the $\Gamma-L_{\rm X}$ correlation}
In Figure \ref{pmao}, we show $\Gamma$ versus $L_{2-10}$ for sources in the
ranges of $0.3 \lesssim z \lesssim 0.96$ (upper panel), and $1.5 \lesssim z
\lesssim 3.3$ (lower panel). The values of the X-ray luminosities and
spectral indices shown in Figure \ref{pmao} were obtained by fitting the
spectra in the observed-frame energy range of 0.5--8~keV (see Table
\ref{tab:CDFs}). We searched for a correlation between $\Gamma$ and $L_{\rm
X}$  by computing the Spearman's  and Kendall's correlations (see Table
\ref{tab:corr}). We find a strong correlation between $\Gamma$ and
$L_{2-10}$, at the $>$99.9\% confidence, for sources having $0.3 \lesssim z
\lesssim 0.96$ and $1.5 \lesssim z \lesssim 3.3$. We tested for a linear
dependence between $\Gamma$ and log~$L_{\rm X}$ by calculating the
Pearson's correlation and find a high significance ($>$99.8\%) for sources
within \fbin~ and \tbin~ (see Table \ref{tab:corr}). In Table
\ref{tab:minl}, we also present results of linear least-squares fits to the
$\Gamma-L_{\rm X}$ relation with a model of the form
$\Gamma=\alpha$~log~$L_{\rm X}+\beta$. For this test, we assumed that
$\Gamma$ is the dependent variable with errors given at the 68\% confidence
level.
In Table \ref{tab:minl}, we show the best-fit linear fit parameters
$\alpha$ and $\beta$. We find that the best-fit parameters $\alpha$ and
$\beta$ show a significant change between the redshift bin $0.3 \lesssim z
\lesssim 0.96$ and the redshift bins of $0.96 \lesssim z \lesssim 1.5$ and
$1.5 \lesssim z \lesssim 3.3$. In particular, for spectral fits performed
in the 0.5 -- 8~keV observed-frame band we find the following: The slope
and offset of the linear fit to the $\Gamma-L_{\rm X}$  correlation in the
$0.3 < z < 0.96$ redshift range are, $\alpha$= 0.14 $\pm$ 0.02 and $\beta =
-4.5 \pm 0.8$, respectively. The slope and offset of the $\Gamma-L_{\rm X}$
correlation in the $1.5 < z < 0.33$ redshift range are, $\alpha$ = 0.23
$\pm$ 0.03,  $\beta = -8.7 \pm 1.2$, respectively. Similar result are found
for spectral fits performed in the 2--10~keV rest-frame. This change in the
linear parameters can also be seen in Figure \ref{alll}.

\input{t5.tex}

\begin{figure}
   \epsscale{1} \plotone{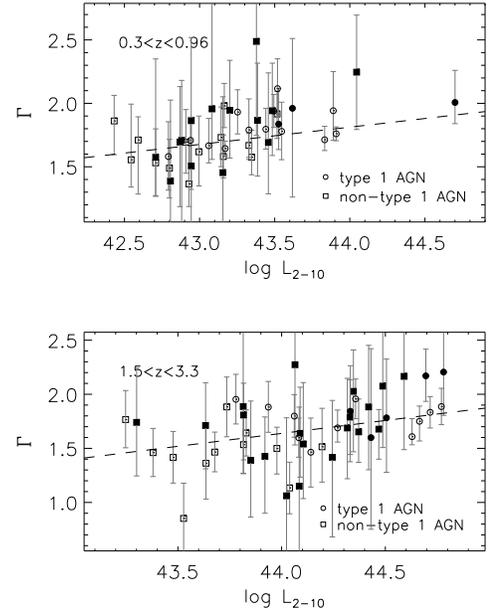}
        \centering
      \caption{$\Gamma$ versus 2--10~keV luminosity of radio-quiet AGNs with
      \fbin~ (upper panel) and with \tbin~ (lower panel).
      The values of the X-ray luminosities and spectral indices were obtained by fitting the spectra in the
      rest-frame energy range of \rest~ (see Table \ref{tab:CDFs}). The dashed lines indicate
      linear fits to the data using the least-squares method. The open symbols correspond
      to sources having $\mbox{\lnh~$\lesssim$~22}$, and the filled symbols are sources with \lnh~$>$~22.
      Circles correspond to type 1 AGNs and squares to non-type 1 AGNs.}
      \label{pmar}
\end{figure}

In Figure \ref{conf} we show the 68\% and 99.9\% confidence contours of
$\alpha$ and $\beta^{*}$ for AGNs in the $0.3 < z < 0.96$ and $1.5 < z <
0.33$ redshift ranges and for fits performed in the 2--10~keV rest-frame
band (Figure \ref{conf}a) and in the 0.5--8~keV observed-frame band (Figure
\ref{conf}b). The parameter $\beta^*$ is obtained from fits of the
model $\Gamma~=~\alpha~ {\rm log~} \frac{L_{2-10}}{(10^{44}~{\rm
ergs/s)}}~+~\beta^*$. $L_{2-10}$ was re-normalized for the purpose of
illustrating better the full range of the contours. The 68\% and 99.9\%
confidence contours levels correspond to $\Delta \chi^2(\alpha,\beta^*)$
values of 2.3 and 13.81, respectively. The confidence contours indicate
that the parameters of the linear fit to the $\Gamma-L_{\rm X}$ correlation
change at the $>$~99.9\% confidence level between the  $0.3 < z < 0.96$ and
$1.5 < z < 0.33$ redshift ranges.

To test the sensitivity and stability of these confidence contours to
possible outliers in the data we repeated the confidence contour analysis
by excluding data points with significant deviations from the linear fit.
In particular, we re-fit the $\Gamma-L_{\rm X}$  correlation and
re-calculated the confidence contours after excluding data points that
deviated by more than $2\sigma$, $2.5\sigma$ and $3\sigma$ from the linear
fit. In all cases we find that the parameters of the linear fit to the
$\Gamma-L_{\rm X}$  correlation change between redshift bins 1 and 3 at the
$>$~99.9\% and $>$~98\%  confidence levels for fits performed in the
0.5--8~keV observed-frame and 2--10~keV rest-frame, respectively.

As discussed in \S4 to test the influence of possible effects such as
Compton reflection, soft excesses, and intrinsic absorption on the $\Gamma$
-- $L_{\rm X}$ correlation, we also fitted the spectra in the \rRF, where
these effects are expected to be smaller. The results of these spectral
fits are presented in Table \ref{tab:CDFs}. In Figure \ref{pmar}, we
present $\Gamma$ versus $L_{2-10}$ for sources in the redshift range of
$0.3 \lesssim z \lesssim 0.96$ (upper panel), and in the redshift range of
$1.5 \lesssim z \lesssim 3.3$ (lower panel) for spectral fits performed in
the 2--10~keV rest-frame band. The results of our correlation analysis
applied to the variables $\Gamma$ and $L_{\rm X}$ are shown in Table
\ref{tab:corr}. We find the Spearman, Kendall and Pearson correlation
coefficients of $\Gamma$ vs. $L_{2-10}$ to be significant at the  $>$99.9\%
and $>$99.7\% confidence levels, for sources within \fbin$ $ and \tbin$ $
respectively. These results suggest that Compton reflection, soft excesses,
and intrinsic absorption are most likely not driving the observed
correlation between $\Gamma$ and $L_{\rm X}$ in the two redshift bins
analyzed in this section. In \S 5.2.3 and \S 5.3.6, we provide detailed
analyses to show that intrinsic absorption and Compton reflection have
negligible contributions to the \Glx correlation.

\begin{figure}
   \epsscale{.8} \plotone{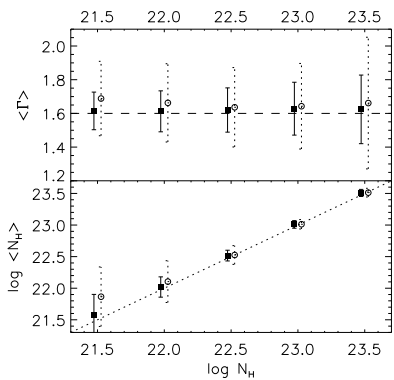}
        \centering
      \caption{Results from fits performed on 1000 simulated spectra with
              $S=550$, $\Gamma=1.6$, $z$=1.4 and 5 different values
              of log~$N_{\rm H}$ (21.5, 22, 22.5, 23 and 23.5). The estimated parameters with fits performed
              in the 0.5--8~keV observed-frame band are shown with filled squares (solid error bars),
              and the estimated parameters with fits performed in the 2--10~keV rest-frame band are
              shown with open circles (dashed error bars). A small shift in the value of \lnh$ $
              (horizontal axis) has been introduced for visual purposes. In the upper panel we show
              $\langle \Gamma \rangle$ vs. \lnh. In the lower panel  we show log$\langle N_{\rm H}\rangle$ vs \lnh.
              All error bars represent $\pm$ 1 $\sigma$ deviations.}
            \label{simu}
\end{figure}

\input{t6.tex}

\input{t7.tex}

\subsubsection{Dependence of the $\Gamma-L_{\rm X}$ correlation on $N_{\rm H}$}

The estimated values of the photon indices used in our correlation analysis
depend partially on the assumed spectral models used to fit the AGN
spectra. In particular, the default model used in our spectral analysis
assumes a simple power law that can be modified by intrinsic absorption.
There is some evidence suggesting that the intrinsic column density
($N_{\rm H}$) could be evolving both with X-ray luminosity
\citep[e.g,][]{Ued03,Aky06} and redshift \citep[e.g.,][]{Aky06,Tre06}. At
the same time, large values of $N_{\rm H}$ could be producing some
dispersion in the estimated values of $\Gamma$. In order to analyze the
effect of $N_{\rm H}$ in the spectral fitting, we have performed
simulations using the software command ``fakeit'' in XSPEC. We randomly
generated 1000 fake spectra for each of five values of $N_{\rm H}$
(\lnh~=~21.5, 22, 22.5, 23 and 23.5). Each simulated spectrum was created
assuming an absorbed power-law (APL) model with 550 counts in the
0.5--8~keV band, $\Gamma=1.6$, and $z=1.4$. The simulated sources were
considered close to the aim-point of the \chandra\ ACIS-I CCD. The assumed
values of the total counts, $\Gamma$ and $z$ are close to the mean values
found in \S4. We performed fits to the randomly generated spectra using the
same APL model, and plot in Figure \ref{simu} (upper panel), the mean
spectral slope (with standard deviation) of the 1000 fits as a function of
\lnh; these fits were performed both in the 0.5--8~keV observed-frame
(squares) and the 2--10~keV rest-frame (circles). Based on these results we
do not find any significant bias in the estimation of $\Gamma$ with $N_{\rm
H}$. We do, however, find that the standard deviation shows a clear
tendency to grow with $N_{\rm H}$ independently of the energy band fit, as
seen in Figure \ref{simu} (upper panel). In Figure \ref{simu} (lower
panel), we see that in general the estimated value of $N_{\rm H}$ is
accurate for \lnh~$\gtrsim 22$; however, for \lnh~ $<$~22 the column
density is slightly overestimated and has a larger dispersion.

Using the same simulations, we estimated the effectiveness of using the
$F$-test at the 95\% level of significance to determine the improvement in
the fit quality by using an absorbed power-law (APL) model as an
alternative to the default power-law (PL) model. Table \ref{tab:simu} shows
that in a simulation of 1000 fake spectra with $S=550$, $\Gamma=1.6$,
$z=1.4$ and \lnh~=~22, the $F$-test indicates absorption in $\sim$75\% of
the spectra. For simulated spectra with \lnh~=~22.5,  the $F$-test
indicates absorption in $\sim 99$\% of the cases and for \lnh~$>$~22.5 the
$F$-test indicates absorption in more than 99.9\% of the cases. Based on
these simulations, we conclude that the $F$-test can accurately identify
absorption when \lnh~$\gtrsim$~22.

Based on our finding that highly-absorbed sources show a greater dispersion
of the estimated value of $\Gamma$, we tested the sensitivity of the
$\Gamma-L_{\rm X}$ correlation for sources having $0.3 \lesssim z \lesssim
0.95$ and $1.5 \lesssim z \lesssim 3.3$ to intrinsic absorption, by
removing sources with significant absorption (\lnh~$\gtrsim$~22.5). We also
tested this correlation for sources having \lnh~$<$~22. Finally as a
complementary test we analyzed the $\Gamma-L_{\rm X}$ correlation for type
1 AGNs. The results of these three tests are presented in Table
\ref{tab:conh}. For sources having \lnh~$\lesssim$~22.5 and
\lnh~$\lesssim$~22 we find in the first and third redshift bins that the
Spearman correlation coefficients of $\Gamma$ versus $L_{2-10}$ are
significant at the $>$95\% confidence levels. This result holds for sources
with fits performed in both the \oOF $ $ and in the 2--10~keV rest-frame
(see Table \ref{tab:conh}). Notice that sources having \lnh~$\lesssim$~22
are plotted as empty squares in Figures \ref{pmao} and \ref{pmar}.

For type 1 AGNs, we find that the  $\Gamma-L_{\rm 2-10}$ correlation is
significant at the 82\% and 12\% levels in the first and third redshift
bins, respectively, for fits performed in the 0.5--8~keV observed-frame
band; the significances are at the 98\% and 72\% levels, respectively, for
fits performed in the 2--10~keV rest-frame band. We briefly investigate
possible reasons that may explain the apparent low detection significance
of the $\Gamma-L_{\rm 2-10}$ relation for the type 1 AGNs of our sample.
First we note that the luminosity ranges of the type~1 AGNs of our sample
in the first and third redshift bins are $42.6 \lesssim \log{L_{2-10}}
\lesssim 44.7$ and $43.8 \lesssim \log{L_{2-10}} \lesssim 44.8$. Our sample
of type 1 AGNs therefore includes relatively luminous sources in each
redshift bin. For sources in the third redshift bin, as we will later see
in \S 5.2.7, the values of $\Gamma$ appear to saturate above
\lLa~$\sim$~45. Therefore, the type 1 AGNs detected in the third redshift
bin of our sample are expected to lie on the flat part of the
$\Gamma-L_{\rm X}$ relation. We conclude that the apparent low significance
of the $\Gamma-L_{\rm 2-10}$ relation for the type 1 AGNs of our sample
found in the third redshift bin is mainly the result of their relatively
large luminosity and the limited number of type 1 AGNs in our sample. Based
on the tests presented in this section we confirm that the strong
$\Gamma-L_{\rm X}$ correlations that we find in RQ AGNs in the redshift
ranges of $0.3 \lesssim z \lesssim 0.95$ and $1.5 \lesssim z \lesssim 3.3$
are not driven by intrinsic absorption.

\begin{figure}
   \epsscale{1} \plotone{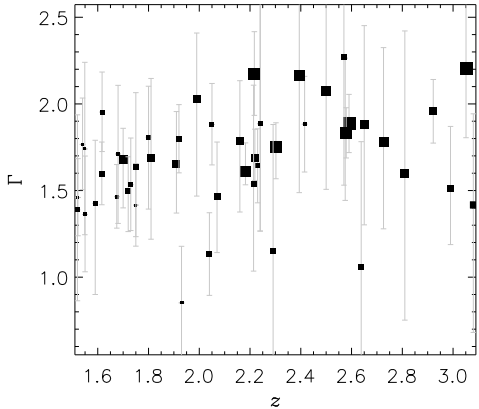}
        \centering
      \caption{Spectral slope ($\Gamma$) versus redshift ($z$) of radio-quiet
       AGNs with $1.5 \lesssim z \lesssim 3.3$ with fits performed in the 2--10~keV rest-frame.
       The size of the symbols increases with $L_{2-10}$. }
        \label{zGam}
\end{figure}

\input{t8.tex}

\subsubsection{Other Correlations}
We searched in the redshift ranges of $0.3 \lesssim z \lesssim 0.95$ and
$1.5 \lesssim z \lesssim 3.3$ for possible correlations between the photon
index $\Gamma$ and other physical parameters of our sample of AGNs by
computing Spearman rank correlations. The results from this correlation
analysis are shown in Table \ref{tab:copa}. In the cases where the
intrinsic column densities were $\mbox{$N_{\rm H}\lesssim 10^{20}~{\rm
cm^{-2}}$}$, we could only obtain upper limits to $N_{\rm H}$, and
therefore computed the correlation coefficients using survival analysis
\citep{Iso86}.

The selection criteria used in this work impose a luminosity limit which is
redshift dependent (see Figure \ref{sele}). Furthermore, the co-moving
density of luminous AGNs is known to increase with $z$. These two effects
will produce a correlation between $L_{\rm X}$ and $z$ as can be seen in
Table \ref{tab:copa}. We do not find any significant correlation between
$\Gamma$ and $N_{\rm H}$ in any bin. The fact that $\Gamma$ is not
correlated with $N_H$ provides further support that the $\Gamma-L_{\rm X}$
correlation is not driven by $N_H$.

We also find a weak correlation between $N_{\rm H}$ and $z$  in the
redshift range of $1.5 \lesssim z \lesssim 3.3$ for fits in the 2--10~keV
rest-frame band. This result may imply that the intrinsic column density
evolves, increasing with redshift. Such a result has been reported in
several studies \citep[e.g.,][]{LaF05,Tre06}; however, the evolution of
$N_{\rm H}$ with $z$ is still a debatable topic since other authors have
not found definitive evidence for the evolution in the ``obscuration
fraction'' \citep[e.g.,][]{Ued03,Aky06,Dwe06}.

A rather surprising result was the detection of a correlation between
$\Gamma$ and $z$ in the third redshift bin with an apparent significance of
99.6\%. A careful analysis indicates that this apparent $\Gamma-z$
correlation is most likely the result of selection effects. This tendency
seems to be confirmed in Figure \ref{zGam}. This $\Gamma$ versus $z$ plot
indicates that higher luminosity sources tend to group in the upper right
area and lower luminosity sources in the lower left area. To test for
selection effects, we performed a correlation analysis including sources
with luminosities greater than the minimum luminosity of a detectable
source at $z \sim 3$. This limit corresponds to \lLa~$\sim$~44 (see Figure
\ref{sele}). We find that the Spearman's correlation probability of the
$\Gamma-z$ correlation in the third redshift bin for fits performed in the
2--10~keV rest-frame, decreases to a non-significant level of $\sim$ 66\%
when we only include sources with \lLa~$\gtrsim$~44 (30 RQ AGNs). We note,
however, that within the same luminosity range the $\Gamma-L_{2-10}$
correlation is significant at the $>$99\% ($r_C \sim 0.6$) level. Our
analysis indicates that the apparent correlation between $\Gamma$ and $z$
in the third redshift bin is most likely the result of selection effects.
This conclusion is also confirmed in \S 5.2.7.

\begin{figure}
   \epsscale{0.8} \plotone{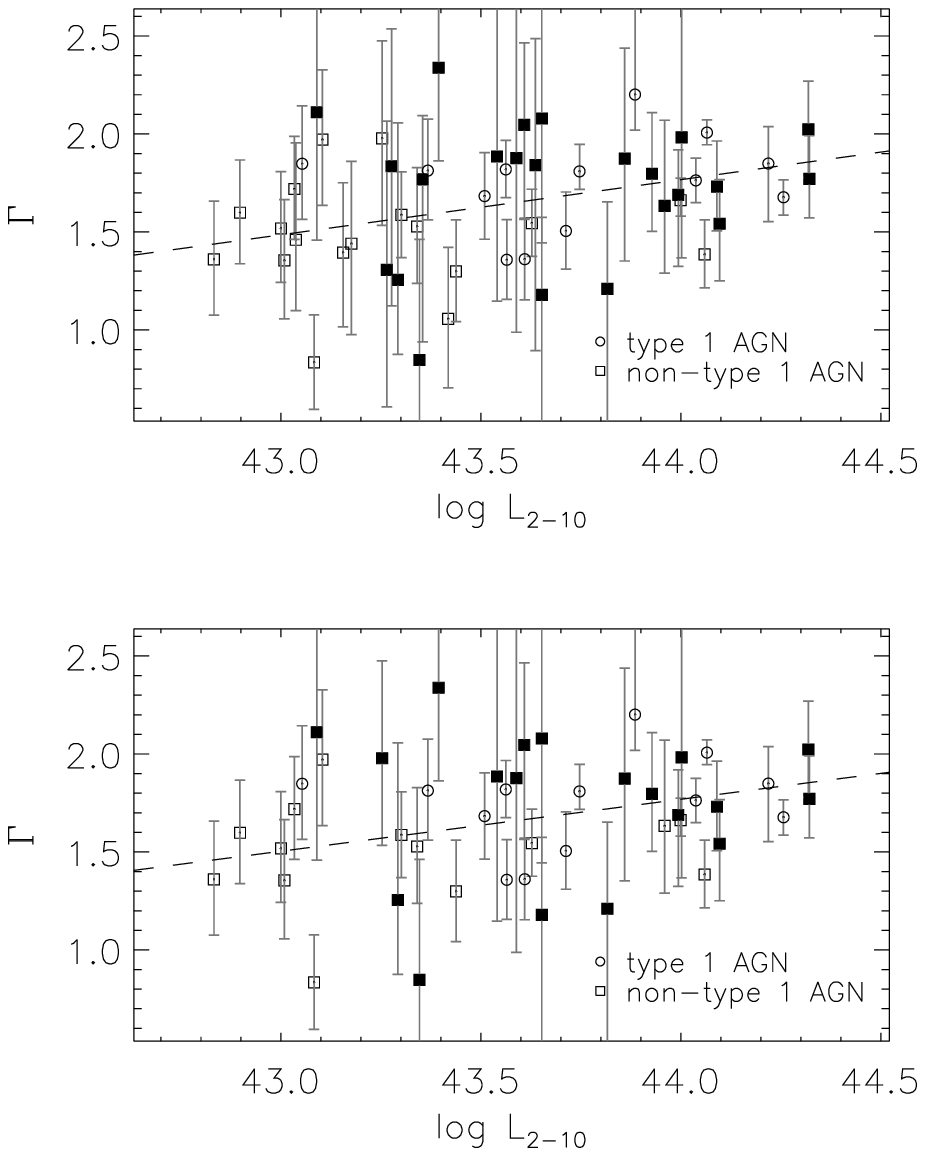}
        \centering
      \caption{$\Gamma$ versus 2--10~keV luminosity of radio-quiet AGNs
 having \sbin $ $ with fits performed in the \oOF $ $ band (upper panel) and fits performed in the \rRF $ $ band (lower panel). The dashed lines indicate linear fits to the data using the least-squares method. The open symbols represent sources having $\mbox{\lnh~$\lesssim$~22}$, and the filled symbols represent sources with \lnh~$>$~22. Circles represent type 1 AGNs and squares non-type 1 AGNs.}
        \label{sbin}
\end{figure}

\subsubsection{Radio-Quiet AGNs with \sbin}

In \S 5.2.1, we showed that the $\Gamma-L_{\rm X}$ correlation was not
significant for sources having $0.96\lesssim z \lesssim 1.5$, especially
for fits performed in the \rRF $ $ where we found that the correlation was
only significant at the $\sim$85\% confidence level. In this section we
investigate the cause of the lower significance of the $\Gamma-L_{\rm X}$
relation for sources having $0.96\lesssim z \lesssim 1.5$. In Figure
\ref{sbin}, we show $\Gamma$ versus $L_{2-10}$ for sources having \sbin $ $
with fits performed in the \oOF $ $ (upper panel) and \rRF $ $ (lower
panel). The $\Gamma$ versus $L_{2-10}$ data points show a larger scatter
than what is seen in the other redshift bins consistent with the lower
significance found for the $\Gamma-L_{\rm X}$ correlation. In
Table~\ref{tab:conh}, we present the results of our correlation analysis of
the $\Gamma-L_{\rm X}$ data for sub-samples of different intrinsic
absorption. We find that sources with \lnh~$\gtrsim$~22 and \sbin $ $ show
no significant correlation between $\Gamma$ and $L_{\rm X}$, whereas
sources with \lnh~$\lesssim$~22 and \sbin $ $ have a $\Gamma-L_{\rm X}$
correlation that is significant at the $ > $90\% confidence level. We
conclude that the absorbed sources with \sbin $ $ are possibly diluting the
correlation significance found in this redshift bin. We caution, however,
that the low number of sources per sub-sample used in this analysis
combined with the uncertainties in the photon indices may result in
inaccurate estimates of the strengths of the correlations.

Our correlation analysis between several other spectral parameters for
sources with \sbin $ $ is included in Table~\ref{tab:copa}. We find an
anticorrelation between $N_{\rm H}$ and $z$ at the $\sim90\%$ confidence
level for sources with \sbin $ $ and for fits performed in the 0.5--8~keV
observed-frame and the 2--10~keV rest-frame bands. Correlations between
$\Gamma$ versus $z$ and $\Gamma$ versus $N_{\rm H}$ for sources in \sbin $
$ are found to be moderately significant for fits performed in the
0.5--8~keV observed-frame band and not significant for fits performed in
the 2--10~keV rest-frame band.

\subsubsection{Dependence of the $\Gamma-L_{\rm X}$ correlation on Compton-reflection}

In this section, we address the possibility that the $\Gamma-L_{\rm X}$
correlation found in this work is produced by a change with luminosity of
the Compton-reflection component. We note that the Compton-reflection
component is difficult to model accurately in low-to-medium S/N X-ray
spectra and therefore inaccurate modeling of this component may result in
apparent flattening of the X-ray spectra.

Several studies indicate that the equivalent width (EW) of the iron
K$\alpha$ emission line in the X-ray spectra of AGNs is anti-correlated
with the 2--10~keV luminosity \citep[e.g.,][]{Iwa93,Nan97,Pag04,Bia07}.
This anti-correlation is commonly referred to as the `X-ray Baldwin effect'
and is also known as the `Iwasawa \& Taniguchi effect' . There are several
proposed physical explanations in the literature for the X-ray Baldwin
effect including (1) a change in the covering factor of a Compton-thick
torus with luminosity \citep[e.g.,][]{KK94}, (2) a luminosity-dependent
ionization state of the iron-emitting material \citep[e.g.,][]{Nan97}, and
(3) variability of the continuum AGN emission assuming constant iron-line
fluxes \citep[e.g.,][]{Jg06}. It has also been proposed that the X-ray
Baldwin effect is driven mostly by changes in the Eddington luminosity
ratio rather than by X-ray luminosity \citep[e.g.,][]{Jg06}.

Several models of AGN accretion disks assume the iron line and
Compton-reflection components originate from X-ray emission reprocessed in
the accretion disk and indicate that the strength of the Compton-reflection
component increases monotonically with the EW of the iron line
\citep[e.g.,][]{Geo91,Ghi94}, and consequently decreases with $L_{\rm X}$
as well. Therefore, under this premise, these models could possibly explain
the $\Gamma-L_{\rm X}$ relation found in this work, since a decrease of the
Compton-reflection component with $L_{\rm X}$ could result in an increase
in $\Gamma$ with $L_{\rm X}$ if the Compton-reflection component is not
modeled accurately in our spectral analysis.

Bianchi et al. (2007) recently found a strong anti-correlation between the
neutral narrow component of the iron K$\alpha$ emission line and the
2--10~keV luminosity of AGNs. These authors suggest that the neutral narrow
iron-line component originates from the molecular torus and the broad
iron-line component originates from reprocessing in the accretion disk. The
dependencies, however, of the Compton-reflection component with the
2--10~keV luminosity are still unclear \citep[e.g.,][]{Nan95,Page04}. It is
also unclear whether the Compton-reflection component observed in the X-ray
spectra of AGNs originates mostly from the torus or the accretion disk. If
the neutral narrow iron K$\alpha$ emission line originates from the torus
and the Compton-reflection component from the accretion disk then one
cannot simply assume that the Compton-reflection component will follow the
X-ray Baldwin effect. Variability studies of individual AGNs such as the
Seyfert~1 galaxies NGC 5548 \citep[]{Chiang00} and MCG-6-30-15
\citep[]{Lee00} indicate that the Compton-reflection component increases
with X-ray luminosity and the iron line EW and the relative normalization
of the Compton-reflection hump are anti-correlated. We note that the
Seyfert~1 galaxies in these variability studies were observed to vary over
a factor of up to $\approx$3 in luminosity whereas our study includes
objects spanning a factor of $\sim$200 in luminosity. It is therefore
difficult to extrapolate the results of these variability studies to our
work.

\begin{figure}
   \epsscale{0.8} \plotone{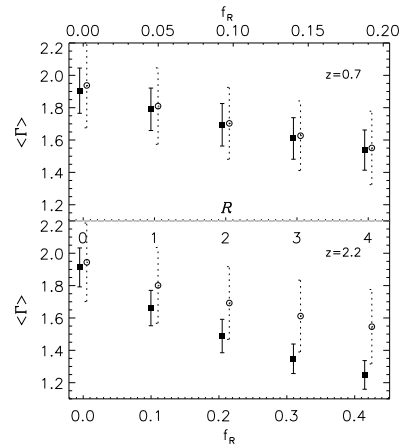}
        \centering
      \caption{Results from fits performed on 1000 simulated spectra with
$S$=550, \mbox{$\Gamma$=1.9}, and 5 different values of the
Compton-reflection scaling factor $R$ ($R$=0,1,2,3,4). The mean of the
best-fit values of $\Gamma$ with fits performed in the 0.5--8~keV
observed-frame band are shown with filled squares (solid error bars), and
the mean of the best-fit values of $\Gamma$ with fits performed in the
2--10~keV rest-frame band are shown with open circles (dashed error bars).
A small shift in the values of $R$ (horizontal axis) has been introduced
for visual purposes. In the upper panel we show $\langle \Gamma \rangle$
vs. $R$ with $z$=0.7. In the lower panel we show $\langle \Gamma \rangle$
vs. $R$ with $z$=2.2. On the x-axis we also show the ratio of photons
($f_R$) in the full band (\obs) that originate from Compton reflection to
photons that originate from the direct power-law component. All error bars
represent $\pm$ 1 $\sigma$ deviations.} \label{simR}
\end{figure}

Since observationally it is still unclear how the Compton-reflection
component depends on X-ray luminosity we have investigated the degree to
which Compton-reflection can be driving the $\Gamma-L_{\rm X}$ relation by
performing simulations and additional tests upon our data. We first
simulated X-ray spectra containing Compton-reflection components with
integer reflection scaling factors ranging between $0 \leq R \leq 4$. For
each value of the reflection scaling factor ($R$=0,1,2,3,4); we simulated
1000 spectra using the FAKEIT command in XSPEC. The Compton-reflection
components were simulated using the PEXRAV model, assuming sources close to
the aimpoint of the ACIS-I CCD, $\Gamma$=1.9, \lnh~=~22,
 a total number of events per spectrum of $S$=550, an e-folding
cutoff energy of $E_{\rm cut}$=400~keV, and an inclination angle ($i$) of
the reflector equal to 30$^{\circ}$. The values of $E_{cut}$ and $i$ were
chosen to be close to those generally used to model Seyfert galaxies
\citep[e.g.,][]{Mag95}. Our results are insensitive to any reasonable value
of $E_{\rm cut}$ and changing $i$ will mostly affect the overall strength
of the reflection component. We performed these simulations assuming
redshifts of $z=0.7$ and $z=2.2$, which correspond to the mean redshifts of
the sources in our sample with \fbin $ $ and \tbin. We proceeded in fitting
the simulated spectra with absorbed power-law models to estimate the
decrease in the fitted values of $\Gamma$ vs. the strength of the
Compton-reflection component. In Figure~\ref{simR} we show the best-fit
values of $\Gamma$ as a function of the reflection scaling factor for
sources with redshifts of $z=0.7$ (upper panel) and $z=2.2$ (lower panel).
In Figure~\ref{simR} we also show the ratio of photons  in the full band
(\obs) that originate from Compton reflection to photons from the direct
power-law component ($f_{\rm R}$).

As expected fits performed in the \rRF $ $ band of the \mbox{high-$z$}
sources are less affected by the Compton-reflection component and show a
smaller change of $\Gamma$ than fits performed in the \oOF $ $ band.
Specifically, we find apparent changes of $\Gamma$ of about 0.7 and 0.3 for
fits performed in the \oOF $ $ and \rRF $ $ bands, respectively, for
sources with \tbin.

Our simulations indicate that if Compton reflection is producing the
observed change in $\Gamma$ of about 0.5 for sources with \tbin $ $ (see
lower panel of Figure \ref{pmao}), then the mean values of $\Gamma$ derived
from fits performed in the \oOF $ $ band should differ by about 0.2 from
the mean values of $\Gamma$ derived from fits performed in the \rRF $ $
band. Our observations indicate that this is not the case. For sources in
the redshift bins of \fbin, \sbin $ $ and \tbin $ $ the differences between
the weighted mean values of $\Gamma$
($\langle\Gamma_{rest}\rangle$-$\langle\Gamma_{obs}\rangle$) obtained from
fits performed in the \oOF $ $ and \rRF $ $ band are $-$0.03$\pm$0.02,
0.02$\pm$0.02 and 0.01$\pm$0.03 (1-$\sigma$ errors), respectively; the
similarity between $\langle\Gamma_{rest}\rangle$ and
$\langle\Gamma_{obs}\rangle$ is consistent with the results found in \S4.
According to our simulations, if the $\Gamma-L_{\rm X}$ correlation were
produced by the Compton-reflection component then these differences in the
weighted mean values of $\Gamma$ would increase with redshift, reaching
values close to 0.2 for sources with \tbin.

We also expect that if Compton reflection is driving the observed
$\Gamma-L_{\rm X}$ relation then the strength and slope of the correlation
for fits performed in the \oOF $ $ band should be significantly stronger
and steeper than the strength and slope of the correlation for fits
performed in the \rRF $ $ band, especially for the high-redshift sources.
This is not the case. As shown in Table \ref{tab:corr}, the strength
of the $\Gamma-L_{\rm X}$ relation in the \oOF $ $ and \rRF $ $ bands for
sources with $1.5 < z < 3.3$ is 0.45 (at the 99.9\% confidence level) and
0.43 (at the 99.8\% confidence level), respectively. From Table
\ref{tab:minl} the ratio of the slopes of the $\Gamma-L_{\rm X}$ relation
for sources with $1.5 < z < 3.3$ is $\alpha_{\rm obs}/\alpha_{\rm
rest}=0.98\pm0.25 $, where $\alpha_{\rm obs}$ and $\alpha_{\rm rest}$ are
the slopes derived from fits performed in the \oOF $ $ and \rRF $ $ bands,
respectively. Based on our simulations the ratio of the slopes of the
$\Gamma-L_{\rm X}$ relation for source with $1.5 < z < 3.3$ would be
approximately $\alpha_{\rm obs}/\alpha_{\rm rest} \sim 1.5$ if
Compton-reflection was driving the correlation.

\begin{figure}
   \epsscale{1} \plotone{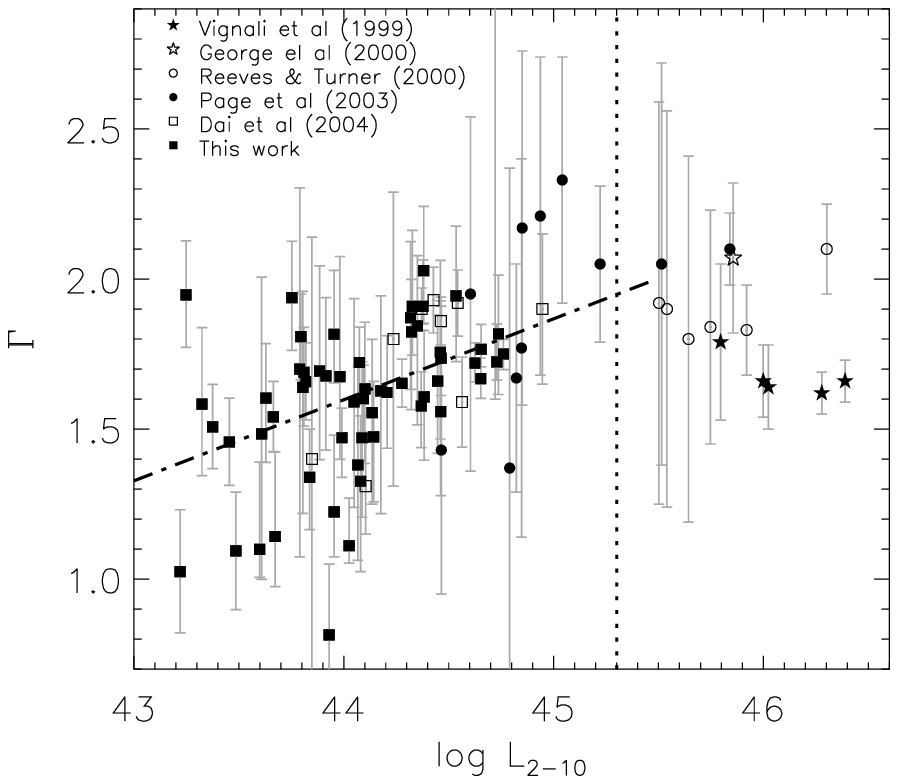}
        \centering
      \caption{Photon indices of \tbin $ $ radio-quiet AGNs obtained from several surveys
versus their 2--10~keV luminosities. Open stars are from \cite{Vig99},
filled stars are from \cite{Geo00}, filled circles are from \cite{Ree00},
open circles are from \cite{Pag03}, open squares are from \cite{Dai04} and
filled squares show data from this work. The vertical dashed line
corresponds to $L_{2-10}=2 \times 10^{45}$ \lumin$ $. The dot-dashed line
shows the best-fit linear model over the luminosity range of $10^{43}-2
\times 10^{45}$ \lumin.}
        \label{Dais}
\end{figure}

\subsubsection{Radio-Quiet AGNs from other surveys with $1.5 \lesssim z \lesssim
3.3$}

We also tested the significance of the $\Gamma-L_{\rm X}$ correlation in
the redshift range of \tbin $ $ by combining results from the independent
surveys of \cite{Vig99}, \cite{Geo00}, \cite{Ree00}, \cite{Pag03}, and
\cite{Dai04} with the results obtained from the CDFs. In Figure \ref{Dais},
we show $\Gamma$ versus $L_{2-10}$ for sources having $1.5\lesssim z
\lesssim 3.3$ combining the results of these surveys with our fits in the
0.5--8~keV observed-frame (Table~\ref{tab:CDFs}). Sources from our survey
fill in the low-luminosity range of the combined data $L_{2-10} \sim
10^{43}-10^{45}$ \lumin. In this range $\Gamma$ increases with $L_X$;
however, for $L_{2-10}\gtrsim 10^{45}$\lumin~it appears that this relation
begins to saturate.

We find that the Pearson linear correlation coefficient reaches a maximum
value for sources with $L_{2-10}$ in the range of $10^{43}-2\times10^{45}$
\lumin, and the Spearman rank coefficient reaches a maximum value for
sources with $L_{2-10}$ in the range of $10^{43}-8\times10^{45}$~\lumin.
There are 76 sources with $L_{2-10}$ in the range of
$10^{43}-2\times10^{45}$ \lumin. The Pearson's correlation coefficient for
\Glx in this luminosity range is $r_p \sim 0.52$ and is significant at the
$>$99.9\% confidence level (null hypothesis probability
$1.8\times10^{-6}$). For reference, the values of the parameters of the
best-fit model of the form $\Gamma=\alpha$~\lLa~+~$\beta$ can be found in
Table~\ref{tab:minl}. There are 84 sources with $L_{2-10}$ in the range of
$10^{43}-8\times10^{45}$ \lumin. The Spearman correlation coefficient in
this luminosity range is $r_S \sim 0.60$ and is significant at the
$>$99.9\% confidence level (null hypothesis probability $\sim
1.6\times10^{-9}$). The limits of the optimized ranges are marked in Figure
\ref{Dais}. Figure \ref{Dais} suggests two different luminosity regimes of
the $\Gamma-L_{\rm X}$ relation. In the first regime that covers the
luminosity range of $L_{2-10} \sim 10^{43}-2\times10^{45}$ \lumin$ $ we
find a linear relation between $\Gamma$ and \lLa. In the second regime,
where $L_{2-10} \gtrsim 2\times10^{45}$ \lumin, we confirm the finding of
\cite{Dai04} that $\Gamma$ decreases with $L_{2-10}$. Specifically, for the
16 sources with $L_{2-10}\gtrsim 2 \times 10^{45}$ \lumin, we found
$\Gamma$ and $L_{\rm X}$ to be anti-correlated with a Spearman correlation
coefficient of $r_S \sim -0.50$ that is significant at the 95.4\%
confidence level.
We also found that for the 84 X-ray luminous sources with $L_{2-10}$ in the
range of $10^{43}-8 \times 10^{45}$ \lumin~ $\Gamma$ and $z$ are not
correlated with a Spearman correlation coefficient of $\sim 0.09$ that is
significant at the $\lesssim$60\% confidence level.

\subsubsection{Physical Interpretation of the $\Gamma-L_{\rm X}$ Relation and its Possible Evolution}

AGN X-ray variability studies have proven very useful for improving our
understanding of the physical structures that produce AGN X-ray spectra.
Several variability studies of individual AGN have found a positive
correlation between $\Gamma$ and the X-ray luminosity,
\citep[e.g.,][]{Mag98,Zdz03}. However, the slope of the $\Gamma-L_{\rm X}$
correlation found in the studies of individual AGNs appears to be
significantly steeper than the $\Gamma$ versus $L_{\rm X}$ slope found in
our current study. For example, \cite{Zdz03} report that a change in
$L_{\rm X}$ by a factor of 10 results in an increase in $\Gamma$ of $\sim$
0.6. For a similar change in $L_{\rm X}$, we find an increase in $\Gamma$
of $\sim$ 0.2--0.3 (see Table \ref{tab:minl}).

To explain the $\Gamma-L_{\rm X}$ correlation, we present two steady state
corona models proposed by \cite{Haa97} and \cite{Mer01}. These coronal
models were originally aimed to explain various X-ray variability
observations of individual objects and therefore assume a constant
black-hole mass. We also introduce a third model \citep{Mer02} focused on
the relative strength of the corona emission and the accretion rate. The
main goal for using this model is to provide a possible connection for the
results obtained in our work with the $\Gamma-L_{\rm bol}/L_{\rm Edd}$
correlations found by \cite{Wan04} and \cite{She06}. At the end of this
section we also comment on how the $\Gamma-L_{\rm X}$ correlation depends
on the optical depth of the hot corona and its evolution with $z$.

The first model posits that the inner accretion disk is sandwiched by a
hot, tenuous and possibly patchy corona \citep{Haa97}. The corona is
coupled to a cooler \mbox{optically-thick} layer (accretion disk), which
provides the seed soft photons that cool the hot layer via inverse Compton
scattering. The spectrum of the scattered photons is in general well fitted
by a power-law and accounts for a large fraction of the observed X-ray
emission in AGNs. This model predicts that $\Gamma$  increases with the
optical depth of the corona $\tau$ and it decreases with the temperature of
the corona. Moreover, if the corona is dominated by $e^\pm$ pairs, the
optical depth of the hot phase is determined by the compactness $\ell$
alone \footnote{$e^\pm$ pair production becomes important for $\ell
\gtrsim 10 $ \cite[e.g., see page 51 of][]{brad97}}, where $\ell$ is
defined as:

\begin{equation} \label{compa}
\ell \equiv \frac{\sigma_T}{m_e c^3} \frac{L_{\rm X}}{R} \approx 10^4 \frac
{\mathcal{L}}{r}
\end{equation}

\noindent Here $\mathcal{L}$ is the luminosity of the corona in Eddington
units, and $r$ is its size in units of Schwarzschild radii. For the case of
a pair-dominated corona this first model predicts that $\Gamma$ will
increase with $L_{\rm X}$. In particular, a change in $L_{\rm X}$ of a
factor of 10 is predicted to produce a $\Delta \Gamma \sim 0.2$. This
predicted variation is consistent with that detected in our sample but
slightly lower than that observed in variability studies of individual
AGNs. When $\mathcal{L} \gtrsim 0.1~r$ ($\ell \gtrsim 1000$) the model
indicates that $\tau$ saturates and does not increase beyond that
luminosity \citep{Haa93}. The first model therefore predicts a flattening
of the $\Gamma-L_{\rm X}$ relation for bright sources as can be seen in
Figure \ref{Dais}.

The second model originally proposed  by \cite{Mer01} assumes a static
patchy corona and is often referred to as the ``thundercloud'' model. The
X-ray spectrum in this model is produced by thermal Comptonization in
spherical regions that are raised above the disc at a given height due to
magnetic flares in active regions of the corona. The thundercloud model
predicts a $\Gamma-L_{\rm X}$ correlation that is consistent with the one
we detect for high-redshift radio-quiet AGNs in the case where the
luminosity of an active region increases with its increasing size at a
given height. Larger active regions tend to be more luminous, cooler and
produce softer spectra (i.e., $\Gamma$ increases). If the size of the
active region gets too large, then a saturation point is reached and
the $\Gamma-L_{\rm X}$ relation becomes flat for luminous sources in
agreement with the combined data sample (Figure \ref{Dais}). The luminosity
of an active region is assumed to scale with its size $r$ via the relation
$L(r)\propto r^D$, where $D$ is a free parameter in the model which may be
related to the internal structure of the region and/or radial dependence of
the energy generation in the accretion disk. Any change in $D$ will
directly affect the slope of the $\Gamma-L_{\rm X}$ relation, making this
model more flexible to explain both our $\Gamma-L_{\rm X}$ correlation in
high redshift AGNs and those found in variability studies of individual AGN
\citep{Zdz03}.

Under the assumption that the correlation is produced by sources of similar
black-hole masses, it seems natural that changes in X-ray luminosity may
result from changes in the Eddington ratio $\epsilon \sim 0.1 \dot{m}$,
where $\dot{m}$ is the accretion rate in units of $\dot{M}_{\rm
Edd}=\frac{L_{\rm Edd}}{c^2}$. Specifically, assuming $L_{\rm X}$ increases
with $L_{\rm bol}$ (see equation 21 of Marconi et al. 2004) and since
\hbox{$L_{\rm bol}= \epsilon L_{\rm Edd}$} we expect, for sources with
similar mass, an increase in $\epsilon$ to result in an increase in $L_{\rm
X}$. A more detailed analysis is provided in \cite{Mer02} in the case of a
$\mbox{coronal-outflow}$ dominated accretion disk model. Under the
assumption that the total power released from the accretion disk-corona
system is $L_{\rm bol} \equiv \epsilon L_{\rm Edd}$, a fraction $f$ will be
released in the corona $L_{\rm X}\approx f \cdot L_{\rm bol}$. Assuming
that magnetic turbulence is the main source of angular momentum transport,
\cite{Mer02} concluded that the relationship between $f$ and $\epsilon$ can
be approximately modeled as a power-law $f\propto \epsilon^{-\delta}$ with
$\delta \sim 0.4$. This relation is mostly independent of the mass of the
black hole $M_{\rm BH}$ \citep{Wan04}. Assuming that $M_{\rm BH}$ is kept
constant then $L_{\rm X} \propto \epsilon^{1-\delta}$, and therefore
$L_{\rm X}$ increases with $\epsilon$. Using these relations in combination
with a steady corona model like the ones already described in this section,
while assuming that $\tau$ increases with $\epsilon$, \cite{Mer02}
concluded that the corona gets cooler and $\Gamma$ increases with
$\epsilon$. This could give a plausible explanation for the correlation
between $\Gamma-L_{\rm bol}/L_{\rm Edd}$ found by \cite{She06}. Based on
the above, and assuming similar black-hole masses, variations in the
Eddington ratio could also explain the correlation found in our work since
the coronal-outflow dominated accretion disk model predicts that $\Gamma$
increases with $L_{\rm X}$.

An alternative explanation is that the $\Gamma-L_{\rm X}$ correlation is
driven by variations in the black-hole masses of the sources and $\epsilon$
is constant, which is in agreement with the predictions of semi-analytic
models by \cite{Kau00}. It is possible that the correlation could be
produced in this case if the optical depth of the corona $\tau$ increases
with the mass of the black hole $M_{\rm BH}$. This is equivalent to
assuming that $\ell$ increases with $M_{\rm BH}$ in the model of
\cite{Haa97} for an $e^\pm$ dominated corona. Since $\mathcal{L} \propto
M_{\rm BH}$, the previous assumption is valid if we assume that the size of
the corona is kept approximately constant as $M_{\rm BH}$ increases (see
equation (\ref{compa})). Under this assumption, both steady corona models
\citep{Haa97,Mer01} analyzed here will reproduce the $\Gamma-L_{\rm X}$
relation.

The possible evolution of the slope and offset of the $\Gamma-L_{\rm X}$
correlation found in this study of RQ AGNs can be explained by an evolution
of the properties of the hot corona. Specifically, the slope of the
$\Gamma-L_{\rm X}$ correlation depends on the optical depth and compactness
parameter of the corona in the model of Haardt et al. (1997) and the
optical depth of the active regions in the thundercloud model by Merloni \&
Fabian (2001). If AGNs within a certain redshift range contain hot coronae
of similar properties we expect them to show a significant correlation
between $\Gamma$ and $L_{\rm X}$ in this redshift range. One explanation of
the possible evolution of the slope and offset of the $\Gamma-L_{\rm X}$
correlation is that the mean properties of the hot coronae of AGN evolve
resulting in a detectable change in the slope and offset of the
$\Gamma-L_{\rm X}$ correlation between AGNs at $ z ~\sim 2.2$ and $ z \sim
0.7 $. One possible explanation for the slight decrease in the strength of
the $\Gamma$ and $L_{\rm X}$ correlation in the second redshift bin is that
this redshift interval is comprised of AGNs having a large range of coronal
properties leading to a weaker correlation between $\Gamma$ and $L_{\rm
X}$. We note that the possible decrease in the significance of the $\Gamma$
and $L_{\rm X}$ correlation in the second redshift bin needs to be
confirmed with a larger sample of RQ AGN.

\section{CONCLUSIONS}

In this paper we have selected a sample of radio-quiet AGNs (173) from the
CDF surveys with moderate-to-high S/N, and have found strong evidence of a
correlation between the X-ray spectral parameters $\Gamma$ and $L_{\rm X}$.
We found that the slope and offset of a linear fit to the $\Gamma-L_{\rm
X}$ correlation possibly evolves for sources with $z \gtrsim 0.1$.
Analyzing this relation in three different redshift bins that contain a
similar number of sources ($\sim$50) we conclude that this correlation is
highly significant in two redshift bins, \fbin, and \tbin $ $ and slightly
less significant in the redshift bin \sbin. We note that the possible
weakness of this correlation for sources with \sbin $ $ appears to be
driven by the absorbed sources in this redshift range. The \Glx
correlations in \fbin $ $ and \tbin $ $ are significant at the $>$99.9\%
confidence level for fits performed in the \oOF $ $ and at the $>$99.5\%
confidence level in the \rRF.

The fact that this correlation is also present when we estimate the
luminosities in the 2--10~keV rest-frame, and also holds for sources with
low column densities, suggests that this correlation is not artificially
driven by any unmodeled complexity in the intrinsic absorption ($N_{\rm
H}$). We performed several tests to investigate whether the \Glx
correlation found in this study is produced by a change with luminosity of
the Compton-reflection component. We found that the strengths and slopes of
the \Glx correlation are similar for fits performed in the 0.5--8~keV
observed-frame and 2--10~keV rest-frame bands. Our analysis indicates that
the strengths and slopes would be significantly different if the
correlation was driven by a Compton-reflection component. The difference
between the observed weighted mean values of $\Gamma$ obtained from fits
performed in the 0.5--8~keV observed-frame and 2--10~keV rest-frame bands
is less than 0.03. Our simulations indicate that if an un-modeled
Compton-reflection component was producing the observed correlation a
difference of $\Gamma$ of about 0.2 would be expected. We conclude that a
Compton-reflection component is unlikely driving the \Glx correlation found
in this study.

This correlation applies to sources with two different luminosity
populations; one with \lLa~$\sim 43.1 \pm 0.5$ (\fbin), and the other with
\lLa~$\sim 44.1 \pm 0.4$ (\tbin) (see Figure \ref{lbin}), indicating
different populations of AGNs. The $\Gamma-L_{\rm X}$ relation results in a
softening of the X-ray spectra as the luminosity of the AGNs increases.

The $\Gamma-L_{\rm X}$ correlation found in the redshift range of $1.5
\lesssim z \lesssim 3.3$ is of special interest because it confirms a
previous independent study of RQQ at $z \gtrsim 1.5$ \citep{Dai04}.
Combining data from \cite{Dai04} and other surveys
\citep{Vig99,Geo00,Ree00,Pag03}, cited in \cite{Dai04}, we find that the
$\Gamma-L_{\rm X}$ correlation becomes even more significant in the
luminosity range of \mbox{$L_{2-10}\sim10^{43}-8\times 10^{45}$\lumin} with
a Spearman correlation coefficient of $r_S \sim 0.6$ significant at the
$>$99.9\% confidence level (null hypothesis probability $\sim
1.6\times10^{-9}$).

We presented two steady-corona models \citep{Haa97,Mer01} that can explain
both the $\Gamma-L_{\rm X}$ correlation found in this work and the
saturation observed in the $L_{2-10}$ vs. $\Gamma$ relation using the
surveys analyzed in \S 5.2.7. Based on these models, we proposed two
different interpretations to explain the $\Gamma-L_{\rm X}$ correlation and
its possible evolution with $z$. The first interpretation posits that this
relation is driven by changes in the Eddington ratio ($\epsilon=L_{\rm
bol}/L_{\rm edd}$) for a population of AGNs of similar mass. The second
interpretation posits that the $\Gamma-L_{\rm X}$ relation is driven by
changes in the mass of the AGNs. The present analysis does not allow us to
infer which of these two scenarios is primarily responsible for driving the
$\Gamma-L_{\rm X}$ correlation; however, future measurements of the
black-hole masses for several AGNs in our sample will allow us to resolve
this issue.

To explain the detected possible evolution of the slope and offset of the
linear fit to the $\Gamma-L_{\rm X}$ correlation we have proposed a simple
model that posits that the mean properties of the hot coronae of AGN at $ z
~\sim 2.2$ differ significantly from those of AGN at $ z \sim 0.7 $. This
model also assumes that within each redshift bin the optical depths of the
hot coronae of the AGNs are similar.

We note that the detected change of the $\Gamma-L_{\rm X}$  correlation
found in our study applies to RQ AGNs detected in the CDFs, which are two
representative and normal fields. Further X-ray spectral studies of deep
\chandra\ fields will test if the detected change of the $\Gamma-L_{\rm X}$
correlation also applies to wider fields of view. Expanding the sample will
also test the correlation in narrower redshift bands and thus better
constrain the epoch at which possible changes in the average emission
properties of AGNs occurred.

\acknowledgments

We would like to thank Michael Eracleous for helpful discussions regarding
the interpretations of our results, and Ohad Shemmer for reviewing the
paper and providing useful comments and suggestions. We also wish to thank
Eric Feigelson and Aaron Steffen for helpful discussions related to several
of the statistical tests implemented in this work. We acknowledge financial
support by NASA grant NAS8-03060. WNB acknowledges financial support from
NASA LTSA grant NAG5-13035.

\end{document}

%% file: t1.tex
\begin{deluxetable}{ccc}
\tabletypesize{\scriptsize}
\tablecolumns{5} \tablewidth{0pt} \tablecaption{Models used in
fitting the spectra of the RQ AGNs of our sample.\tablenotemark{a}
\label{tab:mode}}

\tablehead{ \colhead{~~~~~~Model \tablenotemark{b}~~~~~~} &
\colhead{~~~No sources~~~} & \colhead{\% of the whole sample} }

\startdata

PL & 77 & 44.5   \\
APL & 76 & 43.9    \\
PAPL & 9 &  5.2  \\
IAPL & 4 &  2.3 \\
PL+EL & 4 & 2.3 \\
APL+EL & 3 & 1.7 \\

\enddata
\tablenotetext{a}{The selection criteria for the sample of RQ AGNs
of our present study were that the redshifts of the sources were
greater than 0.1 and the total number of photons in the full band
(0.5--8 keV) was greater than $\sim$ 170. }

\tablenotetext{b}{PL$\equiv$power-law (XSPEC model wabs(pow));
APL$\equiv$absorbed power-law (XSPEC model wabs*zwabs(pow));
PAPL$\equiv$partially absorbed power-law (XSPEC model
wabs*zpcfabs(pow)); IAPL $\equiv$ionized absorbed power-law (XSPEC
model wabs*absori(pow)); PL+EL$\equiv$power-law + emission-line
(XSPEC model wabs(pow+zgauss)); APL+EL$\equiv$absorbed power-law +
emission-line (XSPEC model wabs*zwabs*(pow+zgauss)).}

\end{deluxetable}

%% file: t2.tex
\begin{deluxetable*}{cccccccccc}
\tabletypesize{\scriptsize}
\tablecolumns{10} \tablewidth{0pt} \tablecaption{Properties of our sample
of RQ AGNs selected from the Chandra Deep Field Surveys. \label{tab:CDFs}}

\tablehead{\colhead{Xid \tablenotemark{a}} &
\colhead{$z$ \tablenotemark{b}} & \colhead{Counts \tablenotemark{c}} &
\colhead{$\Gamma$} & \colhead{$N_{\rm H}$\tablenotemark{d}} &
\colhead{log~$L_{2-10}$} & \colhead{$C$-stat} & \colhead{dof} &
\colhead{type\tablenotemark{e}} & \colhead{model \tablenotemark{f}}  }

\startdata
\\
\multicolumn{10}{c}{RESULTS BASED ON FITS PERFORMED IN THE 0.5--8~KEV OBSERVED-FRAME.} \\
\\
\hline
\\

CXOJ123521.32+621628.1 & $0.559^{\rm sp}$ & 513.4 & $1.51^{+0.17}_{-0.07}$ & ..  & 42.79 & 519.2 & 510 & non-type 1 & PL \\
CXOJ123528.77+621427.8 & $0.850^{\rm ph}$ & 183.2 & $1.19^{+0.92}_{-0.42}$ & $4.95^{+4.59}_{-1.97}$ & 43.06 & 542.0 & 509 & non-type 1 & APL \\
CXOJ123529.45+621822.8 & $3.000^{\rm ph}$ & 205.8 & $0.81^{+0.24}_{-0.24}$ & ..  & 43.93 & 516.1 & 510 & non-type 1 & PL \\
CXOJ123535.21+621429.1 & $2.240^{\rm ph}$ & 310.5 & $1.64^{+0.32}_{-0.42}$ & $2.57^{+1.86}_{-2.57}$ & 43.80 & 467.8 & 509 & non-type 1 & APL \\
CXOJ123537.10+621723.6 & $2.050^{\rm sp}$ & 451.1 & $1.82^{+0.21}_{-0.16}$ & ..  & 43.95 & 488.5 & 510 & type 1 & PL \\
CXOJ123539.14+621600.3 & $2.575^{\rm sp}$ & 729.8 & $1.91^{+0.25}_{-0.16}$ & $1.98^{+0.88}_{-1.15}$ & 44.33 & 489.1 & 509 & type 1 & APL \\
CXOJ123546.07+621559.9 & $1.930^{\rm ph}$ & 242.6 & $1.09^{+0.20}_{-0.20}$ & ..  & 43.49 & 474.8 & 510 & non-type 1 & PL \\
CXOJ123548.37+621703.3 & $0.850^{\rm ph}$ & 396.8 & $1.66^{+0.15}_{-0.15}$ & ..  & 42.92 & 440.5 & 510 & non-type 1 & PL \\
CXOJ123548.53+621931.2 & $3.100^{\rm ph}$ & 226.0 & $1.22^{+0.26}_{-0.15}$ & ..  & 43.95 & 439.5 & 510 & non-type 1 & PL \\
CXOJ123550.42+621808.6 & $1.300^{\rm ph}$ & 955.9 & $1.41^{+0.16}_{-0.16}$ & $2.04^{+0.69}_{-0.62}$ & 43.95 & 518.3 & 509 & non-type 1 & APL \\

\\
\hline
\\
\multicolumn{10}{c}{RESULTS BASED ON FITS PERFORMED IN THE 2--10~KEV REST-FRAME.} \\
\\
\hline
\\

CXOJ123521.32+621628.1 & $0.559^{\rm sp}$ & 320.5 & $1.49^{+0.23}_{-0.23}$ & ..  & 42.80 & 386.6 & 349 & non-type 1 & PL \\
CXOJ123535.21+621429.1 & $2.240^{\rm ph}$ & 256.7 & $1.89^{+0.70}_{-0.62}$ & $3.18^{+3.68}_{-3.18}$ & 43.81 & 176.0 & 165 & non-type 1 & APL \\
CXOJ123537.10+621723.6 & $2.050^{\rm sp}$ & 365.7 & $1.88^{+0.24}_{-0.23}$ & ..  & 43.94 & 220.0 & 177 & type 1 & PL \\
CXOJ123539.14+621600.3 & $2.575^{\rm sp}$ & 604.0 & $1.84^{+0.42}_{-0.40}$ & $1.96^{+2.20}_{-1.96}$ & 44.33 & 161.0 & 149 & type 1 & APL \\
CXOJ123546.07+621559.9 & $1.930^{\rm ph}$ & 182.6 & $0.85^{+0.33}_{-0.33}$ & ..  & 43.53 & 218.8 & 184 & non-type 1 & PL \\
CXOJ123548.37+621703.3 & $0.850^{\rm ph}$ & 261.3 & $1.69^{+0.26}_{-0.24}$ & ..  & 42.91 & 268.3 & 293 & non-type 1 & PL \\
CXOJ123550.42+621808.6 & $1.300^{\rm ph}$ & 789.5 & $1.69^{+0.23}_{-0.36}$ & $3.49^{+1.10}_{-1.62}$ & 43.99 & 274.8 & 234 & non-type 1 & APL \\
CXOJ123551.75+621757.1 & $1.910^{\rm ph}$ & 1016.8 & $1.65^{+0.30}_{-0.28}$ & $1.67^{+1.32}_{-1.21}$ & 44.37 & 179.0 & 181 & non-type 1 & APL+EL \\
CXOJ123553.13+621037.3 & $1.379^{\rm sp}$ & 1329.0 & $1.85^{+0.19}_{-0.30}$ & $0.88^{+0.91}_{-0.71}$ & 44.22 & 225.8 & 226 & type 1 & APL \\
CXOJ123555.08+621610.7 & $1.022^{\rm sp}$ & 188.7 & $1.25^{+0.80}_{-0.38}$ & $14.88^{+13.25}_{-9.43}$ & 43.29 & 341.0 & 266 & non-type 1 & PAPL \\

\\

\enddata

\tablenotetext{ }{Note.--- Table 2 is presented in its entirety in the
electronic edition of the Astronomical Journal. A portion is shown here for
guidance regarding its form and content. \\ \\}

\tablenotetext{a}{Xid with RA+DEC coordinates.}
\tablenotetext{b}{Spectroscopic (sp) and photometric (ph) redshifts
gathered from the literature (see \S2).}

\tablenotetext{c}{Background subtracted source counts in the \oOF~ band for
fits performed in the \oOF~ band and counts in the \rRF~ band for fits
performed in the \rRF band.}

\tablenotetext{d}{In units of $10^{22}$ \cmsq.}

\tablenotetext{e}{Based on \cite{Baua04},  source classifications
from \S 4.1.1
(http://www.astro.psu.edu/$\sim$niel/hdf/hdf-chandra.html) }

\tablenotetext{f}{PL$\equiv$power-law (XSPEC model wabs(pow));
APL$\equiv$absorbed-power-law (XSPEC model wabs*zwabs(pow));
PAPL$\equiv$partially-absorbed-power-law (XSPEC model
wabs*zpcfabs(pow)); IAPL $\equiv$ionized-absorbed-power-law (XSPEC
model wabs*absori(pow)); PL+EL$\equiv$power-law + emission-line
(XSPEC model wabs(pow+zgauss)); APL+EL$\equiv$absorbed-power-law +
emission-line (XSPEC model wabs*zwabs*(pow+zgauss)).}

\tablenotetext{g}{The estimated values of $\Gamma$, $N_{\rm H}$,
and X-ray luminosities presented in this table are based on
spectral fits performed in the 0.5--8~keV observed-frame.}

\end{deluxetable*}

%% file: t3.tex
\begin{deluxetable*}{cccccc}
\tabletypesize{\scriptsize}
\tablecolumns{5} \tablewidth{0pt} \tablecaption{Correlation table of
$\Gamma$ versus $L_{\rm X}$. \label{tab:corr}}

\tablehead{ & & & \multicolumn{2}{c}{$\Gamma$ vs. $L_{\rm 2-10}$}  \\
\colhead{Cor. Coeff.} & \colhead{Redshift bin} &
\colhead{Fitted Energy Range} & \colhead{$N$\tablenotemark{a}} &
\colhead{$r_C$} & \colhead{\% sign\tablenotemark{b}}}

\startdata

Spearman & $0.3 \lesssim z \lesssim 0.96$ & \oOF & 53 & 0.48 & $>$99.9  \\
Kendall & $0.3 \lesssim z \lesssim 0.96$ & \oOF & 53 & 0.33 & $>$99.9  \\
Pearson\tablenotemark{c} & $0.3 \lesssim z \lesssim 0.96$ & \oOF & 53 & 0.42 & 99.8 \\
Spearman & $0.96 \lesssim z \lesssim 1.5$ & \oOF & 54 & 0.29 & 96.7 \\
Kendall & $0.96 \lesssim z \lesssim 1.5$ & \oOF & 54 & 0.19 & 95.9 \\
Pearson\tablenotemark{c} & $0.96 \lesssim z \lesssim 1.5$ & \oOF & 54 & 0.31 & 97.5  \\
Spearman & $1.5 \lesssim z \lesssim 3.3$ & \oOF & 57 & 0.45 & $>$99.9  \\
Kendall & $1.5 \lesssim z \lesssim 3.3$ & \oOF & 57 & 0.32 & $>$99.9   \\
Pearson\tablenotemark{c} & $1.5 \lesssim z \lesssim 3.3$ & \oOF & 57 & 0.43 & $>$99.9  \\

Spearman & $0.3 \lesssim z \lesssim 0.96$ & \rRF & 44 & 0.62 & $>$99.9  \\
Kendall & $0.3 \lesssim z \lesssim 0.96$ & \rRF & 44 & 0.42 & $>$99.9 \\
Pearson\tablenotemark{c} & $0.3 \lesssim z \lesssim 0.96$ & \rRF & 44 & 0.530 & $>$99.9 \\
Spearman & $0.96 \lesssim z \lesssim 1.5$ & \rRF & 46 & 0.22 & 84.9  \\
Kendall & $0.96 \lesssim z \lesssim 1.5$ & \rRF & 46 & 0.14 & 83.3 \\
Pearson\tablenotemark{c} & $0.96 \lesssim z \lesssim 1.5$ & \rRF & 46 & 0.23 & 87.8 \\
Spearman & $1.5 \lesssim z \lesssim 3.3$ & \rRF & 48 & 0.43 & 99.8 \\
Kendall & $1.5 \lesssim z \lesssim 3.3$ & \rRF & 48 & 0.31 & 99.8 \\
Pearson\tablenotemark{c} & $1.5 \lesssim z \lesssim 3.3$ & \rRF & 48 & 0.43 & 99.7 \\

\enddata
\tablenotetext{a}{Number of RQ AGNs in each redshift bin. }

\tablenotetext{b}{Percentile significance of the correlation.}

\tablenotetext{c}{Calculated from $\Gamma$ versus log~$L_{\rm
X}$.}

\end{deluxetable*}

%% file: t4.tex
\begin{deluxetable*}{ccccccc}
\tabletypesize{\scriptsize}
\tablecolumns{5} \tablewidth{0pt} \tablecaption{Correlation table of
$\Gamma$ versus $L_{\rm X}$ for AGNs with spectroscopic redshifts.
\label{tab:spec}}

\tablehead{ & & & \multicolumn{2}{c}{$\Gamma$ vs. $L_{\rm 2-10}$} & & \\
\colhead{Fitted Energy Frame} & \colhead{Redshift bin} &
\colhead{$N$\tablenotemark{a}} & \colhead{$r_C$\tablenotemark{b}} &
\colhead{\% sign\tablenotemark{c}} & \colhead{fraction of type 1} &
\colhead{fraction with log$N_{\rm H}\lesssim$22}}

\startdata

\oOF & $0.3 \lesssim z \lesssim 0.96$ & 46 & 0.47 & 99.9  & 0.37 & 0.70\\
\rRF & $0.3 \lesssim z \lesssim 0.96$ & 40 & 0.64 & $>$99.9 & 0.40 & 0.68\\
\oOF & $0.96 \lesssim z \lesssim 1.5$ & 31 & 0.40 & 97.3 & 0.42 & 0.71 \\
\rRF & $0.96 \lesssim z \lesssim 1.5$ & 26 & 0.42 & 96.6 & 0.50 & 0.73\\
\oOF & $1.5 \lesssim z \lesssim 3.3$ & 26 & 0.38 & 94.3 & 0.62 & 0.62 \\
\rRF & $1.5 \lesssim z \lesssim 3.3$ & 24 & 0.49 & 98.5 & 0.67 & 0.67\\

\enddata
\tablenotetext{a}{Number of RQ AGNs in each redshift bin. }

\tablenotetext{b}{Spearman correlation coefficient. }

\tablenotetext{c}{Percentile significance of the correlation.}

\end{deluxetable*}

%% file: t5.tex
\begin{deluxetable*}{ccccc}
\tabletypesize{\scriptsize}
\tablecolumns{5} \tablewidth{0pt} \tablecaption{Results of linear fits to
the $\Gamma$ vs. \lLa~relation. \tablenotemark{a} \label{tab:minl}}

\tablehead{ \colhead{Redshift bin} & \colhead{Fitted energy range} &
\colhead{Sample\tablenotemark{b}} & \colhead{$\alpha$} & \colhead{$\beta$}}

\startdata

$0.3 \lesssim z \lesssim 0.96$ & observed-frame 0.5--8~keV  & CDFs & $0.14\pm0.02$ & $-4.5\pm0.8$  \\

$0.3 \lesssim z \lesssim 0.96$ & rest-frame 2--10~keV   & CDFs  & $0.13\pm0.04$ & $-4.1\pm1.7$  \\

$0.96 \lesssim z \lesssim 1.5$ & observed-frame 0.5--8~keV  & CDFs & $0.23\pm0.03$ & $-8.3\pm1.5$ \\

$0.96 \lesssim z \lesssim 1.5$ & rest-frame 2--10~keV   & CDFs  & $0.27\pm0.05$  & $-9.9\pm2.2 $  \\

$1.5 \lesssim z \lesssim 3.3$ & observed-frame 0.5--8~keV  & CDFs  & $0.23\pm0.03$ & $-8.7\pm1.2$ \\

$1.5 \lesssim z \lesssim 3.3$ & rest-frame 2--10~keV   & CDFs  & $0.24\pm0.06$  & $-8.9\pm2.4 $ \\

$1.5 \lesssim z \lesssim 3.3$ & observed-frame 0.5--8~keV & combined & $0.27\pm0.03$ & $-10.3\pm1.2$ \\

\enddata

\tablenotetext{a}{Based on fits of a linear model ($\Gamma$ =
$\alpha$~\lLa~+~$\beta$) to the $\Gamma$ versus \lLa~relation, using the
weighted least-squares method. The errors in $\Gamma$ at the 68\% level are
used in the linear fit. $L_{2-10}$ is the 2--10~keV rest-frame luminosity.
}

\tablenotetext{b}{The CDFs sample consists of all the sources
presented in Table 2. The combined sample consists of sources
obtained from the independent surveys of \cite{Vig99},
\cite{Geo00}, \cite{Ree00}, \cite{Pag03} and \cite{Dai04} combined
with the sources of our CDFs sample.}

\end{deluxetable*}

%% file: t6.tex
\begin{deluxetable}{ccc}
\tabletypesize{\scriptsize}
\tablecolumns{4} \tablewidth{0pt} \tablecaption{Results of
simulations to test the probability of detecting intrinsic
absorption through spectral fits. \tablenotemark{a}
\label{tab:simu}}

\tablehead{ \colhead{\lnh}  & \hspace{50pt} &
\colhead{{Percentage} of cases{}$^{b}$}}

\startdata

21.5 & &  41.8 \%   \\
22.0 & &  74.7 \%   \\
22.5 & &  99.1 \%  \\
23.0 & &  100.0 \%  \\
23.5 & &  99.9 \% \\

\enddata
\tablenotetext{a}{For each of the 5 different values of \lnh$ $
(\lnh~=~21.5, 22, 22.5, 23 and 23.5), we randomly generated 1000
fake spectra assuming sources close to the aim-point of the
Chandra ACIS-I CCD, with 550 counts in the 0.5--8~keV band,
$\Gamma=1.6$, and $z=1.4$. }

\tablenotetext{b}{Percentage of cases out of the 1000 simulated cases where
the $F$-test indicates a significant presence of absorption assuming the
input values of column densities listed in the first column. }
\end{deluxetable}

%% file: t7.tex
\begin{deluxetable*}{cccccc}
\tabletypesize{\scriptsize}
\tablecolumns{6} \tablewidth{0pt} \tablecaption{Correlation table
of $\Gamma$ vs. $L_{\rm X}$ for sub-samples of different
absorption.   \label{tab:conh}}

\tablehead{ & & & & \multicolumn{2}{c}{$L_{\rm 2-10}$ vs. $\Gamma$}   \\
\colhead{Redshift bin} & \colhead{Absorption{}$^{a}$} &
\colhead{Fitted energy range{}} & \colhead{$N${}$^{b}$} &
\colhead{$r_C${}$^{c}$} & \colhead{\% sign\tablenotemark{d}} }

\startdata

$0.3 \lesssim z \lesssim 0.96$ & log~$N_{\rm H} \lesssim 22.5$ & \oOF & 40 & 0.38 & 98.4 \\
$0.3 \lesssim z \lesssim 0.96$ & log~$N_{\rm H} \lesssim 22.0$ & \oOF & 37 & 0.38 & 98.1  \\
$0.3 \lesssim z \lesssim 0.96$ & log~$N_{\rm H} \gtrsim 22.0$ & \oOF & 16 & 0.66 & 99.4  \\
$0.3 \lesssim z \lesssim 0.96$ & type 1 & \oOF & 17 & 0.34 & 81.6  \\
$0.3 \lesssim z \lesssim 0.96$ & log~$N_{\rm H} \lesssim 22.5$ & \rRF & 32 & 0.56 & $>$99.9 \\
$0.3 \lesssim z \lesssim 0.96$ & log~$N_{\rm H} \lesssim 22.0$ & \rRF & 27 & 0.56 & 99.8 \\
$0.3 \lesssim z \lesssim 0.96$ & log~$N_{\rm H} \gtrsim 22.0$ & \rRF & 17 & 0.67 & 99.7 \\
$0.3 \lesssim z \lesssim 0.96$ & type 1 & \rRF & 16 & 0.58 & 98.2 \\
$0.96 \lesssim z \lesssim 1.5$ & log~$N_{\rm H} \lesssim 22.5$ & \oOF & 38 & 0.27 & 89.9 \\
$0.96 \lesssim z \lesssim 1.5$ & log~$N_{\rm H} \lesssim 22.0$ & \oOF  & 31 & 0.31 & 91.6  \\
$0.96 \lesssim z \lesssim 1.5$ & log~$N_{\rm H} \gtrsim 22.0$ & \oOF  & 23 & 0.31 & 85.4  \\
$0.96 \lesssim z \lesssim 1.5$ & type 1 & \oOF & 17 & 0.20 & 49.5  \\
$0.96 \lesssim z \lesssim 1.5$ & log~$N_{\rm H} \lesssim 22.5$ & \rRF & 33 & 0.26 & 85.3 \\
$0.96 \lesssim z \lesssim 1.5$ & log~$N_{\rm H} \lesssim 22.0$ & \rRF & 27 & 0.33 & 91.1 \\
$0.96 \lesssim z \lesssim 1.5$ & log~$N_{\rm H} \gtrsim 22.0$ & \rRF & 19 & -0.18 & 54.8 \\
$0.96 \lesssim z \lesssim 1.5$ & type 1 & \rRF & 16 & 0.15 & 37.1 \\
$1.5 \lesssim z \lesssim 3.3$ & log~$N_{\rm H} \lesssim 22.5$ & \oOF & 40 & 0.45 & 99.7 \\
$1.5 \lesssim z \lesssim 3.3$ & log~$N_{\rm H} \lesssim 22.0$ & \oOF & 32 & 0.35 & 95.2 \\
$1.5 \lesssim z \lesssim 3.3$ & log~$N_{\rm H} \gtrsim 22.0$ & \oOF & 25 & 0.38 & 94.2 \\
$1.5 \lesssim z \lesssim 3.3$ & type 1  & \oOF & 16 & 0.12 &  34.0 \\
$1.5 \lesssim z \lesssim 3.3$ & log~$N_{\rm H} \lesssim 22.5$ & \rRF & 34 & 0.45 & 99.2 \\
$1.5 \lesssim z \lesssim 3.3$ & log~$N_{\rm H} \lesssim 22.0$ & \rRF & 23 & 0.42 & 95.6 \\
$1.5 \lesssim z \lesssim 3.3$ & log~$N_{\rm H} \gtrsim 22.0$ & \rRF & 23 & 0.42 & 96.2 \\
$1.5 \lesssim z \lesssim 3.3$ & type 1 & \rRF & 16 & 0.29 & 72.1 \\

\enddata

\tablenotetext{a}{Sub-samples contain RQ AGNs that are either type
1 AGN or have $N_{H}$ less than a specified value.}

\tablenotetext{b}{Number of RQ AGNs in each sub-sample. }

\tablenotetext{c}{The Spearman correlation coefficient.}

\tablenotetext{d}{The significance of the Spearman correlation
coefficient.}

\end{deluxetable*}

%% file: t8.tex
\begin{deluxetable*}{cccccc}
\tabletypesize{\scriptsize}
\tablecolumns{5} \tablewidth{0pt} \tablecaption{Correlation table
of $L_{2-10}$ vs. $z$, $\Gamma$ vs. $z$, $N_{\rm H}$ vs. $z$,
$N_{\rm H}$ vs. $L_{2-10}$ and $\Gamma$ vs. $N_{\rm H}$.
\label{tab:copa}}

\tablehead{  \colhead{correlated parameters} &
\colhead{$N$\tablenotemark{a}} & \colhead{Redshift bin} &
\colhead {Fitted energy Range} & \colhead{$r_{\rm c}${}$^{b}$} &
\colhead{\% sign{}$^{c}$} }

\startdata

$L_{2-10}$ vs $z$ & 53 & $0.3 \lesssim z \lesssim 0.96$ & \oOF & $~~0.26$ & 93.5   \\
$\Gamma$ vs $z$ & 53 & $0.3 \lesssim z \lesssim 0.96$ & \oOF & $~~0.12$ & 61.4   \\
$N_{\rm H}$ vs $z$ & 27 & $0.3 \lesssim z \lesssim 0.96$ & \oOF & $-0.08$ & 30.9    \\
$N_{\rm H}$ vs $L_{2-10}$ & 27 & $0.3 \lesssim z \lesssim 0.96$ & \oOF & $~~0.13$ & 50.7   \\
$\Gamma$ vs $N_{\rm H}$ & 27 & $0.3 \lesssim z \lesssim 0.96$ & \oOF & $~~0.14$ & 53.6  \\
$L_{2-10}$ vs $z$ & 44 & $0.3 \lesssim z \lesssim 0.96$ & \rRF & $~~0.29$ & 94.4 \\
$\Gamma$ vs $z$ & 44 &$0.3 \lesssim z \lesssim 0.96$ & \rRF &  $~~0.23$ & 87.4  \\
$N_{\rm H}$ vs $z$  & 25 &$0.3 \lesssim z \lesssim 0.96$ & \rRF & $-0.05$ & 20.6 \\
$N_{\rm H}$ vs $L_{2-10}$  & 25 & $0.3 \lesssim z \lesssim 0.96$ & \rRF & $~~0.08$ & 30.2 \\
$\Gamma$ vs $N_{\rm H}$  & 25 & $0.3 \lesssim z \lesssim 0.96$ & \rRF & $~~0.32$ & 88.5 \\

$L_{2-10}$ vs $z$ & 54 & $0.96 \lesssim z \lesssim 1.5$ & \oOF & $~~0.16$ & 75.6   \\
$\Gamma$ vs $z$ & 54 & $0.96 \lesssim z \lesssim 1.5$ & \oOF & $-0.24$ & 91.8   \\
$N_{\rm H}$ vs $z$ & 28 & $0.96 \lesssim z \lesssim 1.5$ & \oOF & $-0.33$ & 91.8  \\
$N_{\rm H}$ vs $L_{2-10}$ & 28 & $0.96 \lesssim z \lesssim 1.5$ & \oOF & $-0.15$ & 55.4   \\
$\Gamma$ vs $N_{\rm H}$ & 28 & $0.96 \lesssim z \lesssim 1.5$ & \oOF & $-0.35$ & 92.7  \\
$L_{2-10}$ vs $z$ & 46 & $0.96 \lesssim z \lesssim 1.5$ & \rRF & $~~0.17$ & 74.6 \\
$\Gamma$ vs $z$ & 46 & $0.96 \lesssim z \lesssim 1.5$ & \rRF &  $-0.08$ & 41.1  \\
$N_{\rm H}$ vs $z$  & 24 & $0.96 \lesssim z \lesssim 1.5$ & \rRF & $-0.36$ & 91.3 \\
$N_{\rm H}$ vs $L_{2-10}$  & 24 & $0.96 \lesssim z \lesssim 1.5$ & \rRF & $-0.10$ & 36.0 \\
$\Gamma$ vs $N_{\rm H}$  & 24 & $0.96 \lesssim z \lesssim 1.5$ & \rRF & $-0.03$ & 11.8 \\

$L_{2-10}$ vs $z$ & 57 & $1.5 \lesssim z \lesssim 3.3$ & \oOF & $~~0.58$ & $>$99.9  \\
$\Gamma$ vs $z$ & 57 & $1.5 \lesssim z \lesssim 3.3$ & \oOF & $~~0.14$ & 68.4 \\
$N_{\rm H}$ vs $z$ & 34 & $1.5 \lesssim z \lesssim 3.3$ & \oOF & $~~0.13$ & 53.5  \\
$N_{\rm H}$ vs $L_{2-10}$  & 34 & $1.5 \lesssim z \lesssim 3.3$ & \oOF & $-0.23$ & 81.8 \\
$\Gamma$ vs $N_{\rm H}$  & 34 & $1.5 \lesssim z \lesssim 3.3$ & \oOF & $-0.11$ & 48.3 \\
$L_{2-10}$ vs $z$ & 48 & $1.5 \lesssim z \lesssim 3.3$ & \rRF & $~~0.65$ & $>$99.9  \\
$\Gamma$ vs $z$ & 48 & $1.5 \lesssim z \lesssim 3.3$ & \rRF &  $~~0.35$ & 98.6  \\
$N_{\rm H}$ vs $z$ & 30 & $1.5 \lesssim z \lesssim 3.3$ & \rRF & $~~0.34$ & 93.4  \\
$N_{\rm H}$ vs $L_{2-10}$  & 30 & $1.5 \lesssim z \lesssim 3.3$ & \rRF & $-0.02$ & 8.8 \\
$\Gamma$ vs $N_{\rm H}$  & 30 & $1.5 \lesssim z \lesssim 3.3$ & \rRF & $-0.1$ & 40.4 \\

\enddata
\tablenotetext{a}{Number of RQ AGNs in each sub-sample.}

\tablenotetext{b}{The Spearman correlation coefficient.}

\tablenotetext{c}{The significance of the Spearman correlation
coefficient.}

\end{deluxetable*}